\documentclass[twocolumn]{aastex63}
\received{April 13, 2021}
\accepted{July 1, 2021}
\submitjournal{ApJ}

\usepackage{amsmath}

\shorttitle{Multiple merging in LAB1}
\shortauthors{Umehata et al.}
\graphicspath{{./}{figures/}}

\begin{document}

\title{ALMA Observations of Lyman-$\alpha$ Blob 1: Multiple major-mergers and widely distributed interstellar media}

\correspondingauthor{Hideki Umehata}
\email{hideki.umehata@riken.jp}

\author[0000-0003-1937-0573]{Hideki Umehata}
\affil{RIKEN Cluster for Pioneering Research, 2-1 Hirosawa, Wako-shi, Saitama 351-0198, Japan}
\affil{Institute of Astronomy, Graduate School of Science, The University of Tokyo, 2-21-1 Osawa, Mitaka, Tokyo 181-0015, Japan}
\author{Ian Smail}
\affil{Centre for Extragalactic Astronomy, Department of Physics, Durham University, South Road, Durham DH1 3LE, UK}
\author{Charles C. Steidel}
\affil{Cahill Center for Astrophysics, California Institute of Technology, MC 249-17, 1200 East California Boulevard, Pasadena, CA 91125, USA
}
\author{Matthew Hayes}
\affil{Department of Astronomy, Stockholm University, AlbaNova University Centre, SE-106 91, Stockholm, Sweden}
\author{Douglas Scott}
\affil{Department of Physics and Astronomy, University of British Columbia, 6224 Agricultural Road, Vancouver, BC V6T 1Z1, Canada.}
\author{A. M. Swinbank}
\affil{Centre for Extragalactic Astronomy, Department of Physics, Durham University, South Road, Durham DH1 3LE, UK}
\author{R.J. Ivison}
\affil{European Southern Observatory, Karl-Schwarzschild-Str. 2, D-85748 Garching, Germany}
\author{Toru Nagao}
\affil{Research Center for Space and Cosmic Evolution, Ehime University, Bunkyo-cho 2-5, Matsuyama 790-8577, Japan}
\author{Mariko Kubo}
\affil{Research Center for Space and Cosmic Evolution, Ehime University, Bunkyo-cho 2-5, Matsuyama 790-8577, Japan}
\author[0000-0002-6939-0372]{Kouichiro Nakanishi}
\affiliation{National Astronomical Observatory of Japan, 2-21-1 Osawa, Mitaka, Tokyo 181-8588, Japan}
\affiliation{Department of Astronomy, School of Science, The Graduate University for Advanced Studies (SOKENDAI), 2-21-1 Osawa, Mitaka, Tokyo, 181-8588 Japan}
\author{Yuichi Matsuda}
\affiliation{National Astronomical Observatory of Japan, 2-21-1 Osawa, Mitaka, Tokyo 181-8588, Japan}
\affiliation{Department of Astronomy, School of Science, The Graduate University for Advanced Studies (SOKENDAI), 2-21-1 Osawa, Mitaka, Tokyo, 181-8588 Japan}
\author{Soh Ikarashi}
\affil{Centre for Extragalactic Astronomy, Department of Physics, Durham University, South Road, Durham DH1 3LE, UK}
\author{Yoichi Tamura}
\affil{Division of Particle and Astrophysical Science, Graduate School of Science, Nagoya University, Aichi 464-8602, Japan}
\author{J. E. Geach}
\affil{Centre for Astrophysics Research, University of Hertfordshire, Hatfield, AL10 9AB, UK}





\begin{abstract}

We present observations of a giant Lyman-$\alpha$ (Ly\,$\alpha$) blob in the SSA22 proto-cluster at $z=3.1$, SSA22-LAB1, taken with the Atacama Large Millimeter/submillimeter Array (ALMA). 
Dust continuum, along with [C\,\textsc{ii}]\,158$\mu$m, and CO(4--3) line emission have been detected in LAB1, showing complex morphology and kinematics across a $\sim100$\,kpc central  region.
Seven galaxies at $z=$3.0987--3.1016 in the surroundings are identified in [C\,\textsc{ii}] and dust continuum emission, with two of them potential companions or tidal structures associated with the most massive galaxies.
Spatially resolved [C\,\textsc{ii}] and infrared luminosity ratios for the widely distributed media ($L_{\rm [CII]}/L_{\rm IR}\approx10^{-2}-10^{-3}$) suggest that the observed extended interstellar media are likely to have originated from star-formation activity and the contribution from shocked gas is probably not dominant.
LAB1 is found to harbour a total molecular gas mass $M_{\rm mol}=(8.7\pm2.0) \times 10^{10}$\,M$_\odot$, concentrated in the core region of the Ly\,$\alpha$-emitting area.
While (primarily obscured) star-formation activity in the LAB1 core is one of the most plausible power sources for the Ly\,$\alpha$ emission, multiple major-mergers found in the core may also play a role in making LAB1 exceptionally bright and extended in Ly\,$\alpha$ as a result of cooling radiation induced by gravitational interactions.
\end{abstract}



\section{Introduction} \label{sec:intro}

In recent decades, bright and extended nebulae emitting H\,\textsc{i}~Ly\,$\alpha$ emission have been identified in the early Universe. These nebulae, which have extents of several tens to several hundred (physical) kpc and Ly\,$\alpha$ luminosity $L_{\rm Ly\alpha}\sim10^{43}-10^{45}$\,erg\,s$^{-1}$, are called Ly\,$\alpha$ blobs (LABs, e.g., \citealt{1996ApJ...457..490F};
\citealt{1998MNRAS.298..583I};
\citealt{1999AJ....118.2547K}; \citealt{2000ApJ...532..170S};
\citealt{2004AJ....128..569M}; \citealt{2005ApJ...629..654D}; \citealt{2009ApJ...696.1164O}; \citealt{2009ApJ...693.1579Y}; \citealt{2011MNRAS.410L..13M};
\citealt{2014Natur.506...63C};
\citealt{2015Sci...348..779H};
\citealt{2017ApJ...837...71C};
\citealt{2019PASJ...71L...2K}).

The extended emission suggests the presence of plentiful hydrogen on circumgalactic medium (CGM) scales and provides clues to understanding galaxy formation and evolution. For instance, the relation of LABs to the formation of massive galaxies has been proposed (e.g., \citealt{2005ApJ...629..654D}; \citealt{2006ApJ...640L.123M}) and some works suggest that LABs preferentially reside in proto-clusters (e.g., \citealt{2004AJ....128..569M}; \citealt{2018PASJ...70S..14S}). Recently \citet{2019Sci...366...97U} discovered Ly\,$\alpha$ filaments on $\gtrsim 1$ physical Mpc scales in the $z=3.1$ SSA22 proto-cluster. The filaments encompass two LABs reported in \citet{2004AJ....128..569M}, which demonstrates that LABs may be bright knots within gas filaments that are extended over much larger scales and provide fuel for galaxy growth.

What mechanisms produce the Ly\,$\alpha$ emission is also a subject of debate. The scenarios proposed so far include: gravitational cooling radiation associated with pristine, cool hydrogen gas flow (e.g., \citealt{2009MNRAS.400.1109D}; \citealt{2009ApJ...703.1416F}); galactic winds from starbursts (\citealt{2000ApJ...532L..13T}); and photo-ionization driven by star-forming galaxies or active galactic nuclei (AGNs), followed by  scattering  (e.g., \citealt{2009ApJ...700....1G}; \citealt{2011Natur.476..304H}; \citealt{2011ApJ...736..160S}).

The SSA22 proto-cluster at $z=3.1$ is known to harbour several LABs and hence provides a unique laboratory (\citealt{2004AJ....128..569M}). \citet{2011MNRAS.410L..13M} performed a 2.1\,deg panoramic survey to discover fourteen LABs with linear extents over 100\,kpc. Interestingly, the largest and brightest LAB is one of the first discovered LABs, SSA22-LAB1 located close to the proto-cluster core (hereafter LAB1 in this paper, \citealt{2000ApJ...532..170S}). LAB1 has an extent of $\approx$200\,kpc and a luminosity $L_{\rm Ly\alpha}=1.1\times10^{44}$\,erg\,s$^{-1}$ (\citealt{2004AJ....128..569M}), making LAB1 one of the most spectacular LABs known to date. Together with its environment, a remarkable proto-cluster, LAB1 has been intensively investigated by a number of works (e.g.,
\citealt{2001ApJ...548L..17C}; \citeyear{2004ApJ...606...85C}; 
\citealt{2004MNRAS.351...63B}; \citealt{2005MNRAS.363.1398G}; \citeyear{2009ApJ...700....1G}; \citeyear{2014ApJ...793...22G}; \citeyear{2016ApJ...832...37G}; \citealt{2007ApJ...667..667M}; \citealt{2010MNRAS.402.2245W}; \citealt{2011Natur.476..304H};  \citealt{2012ApJ...750..116U}; \citealt{2013MNRAS.430.2768T}; \citealt{2016MNRAS.455.3333K}; \citealt{2016MNRAS.460.4075H};  \citealt{2017ApJ...834L..16U}; \citealt{2017ApJ...850..178A}; \citealt{2020A&A...642A..55H}; \citealt{2021MNRAS.502.2389L}).

For a comprehensive understanding of the nature of LABs and their role in galaxy formation and evolution, observations at (sub)mm wavelengths are of huge importance. Massive star-forming galaxies are easily enshrouded by dust in an intensely star-forming phase, and often undetectable in the optical to NIR (e.g., \citealt{2020A&A...640L...8U} and references therein). Furthermore, the molecular/fine-structure lines at these wavelengths provide powerful tools to characterize the nature and conditions of the interstellar medium (ISMs) in galaxies.
Following a number of attempts since the discovery (e.g., \citealt{2001ApJ...548L..17C}; \citeyear{2004ApJ...606...85C}; \citealt{2005MNRAS.363.1398G}; \citeyear{2014ApJ...793...22G}; \citeyear{2016ApJ...832...37G}; \citealt{2007ApJ...667..667M}; \citealt{2013MNRAS.430.2768T}), the advent of ALMA allows us to revolutionize our picture of LAB1 in this regard. \citet{2016ApJ...832...37G} identified three dusty star-forming galaxies toward LAB1 with a total star-formation rate (SFR) of $\sim200$\,M$_\odot$\,yr$^{-1}$, although only one of the galaxies had a reliable spectroscopic redshift. They proposed that these galaxies identified by ALMA are the dominant sources that power the Ly\,$\alpha$ emission.

Recently \citet{2017ApJ...834L..16U} detected the  [C\,\textsc{ii}]\,158\,$\mu$m emission line from one massive, dusty star-forming galaxy in LAB1. [C\,\textsc{ii}]~158\,$\mu$m ($^2P_{3/2}\rightarrow^2P_{1/2}$) is the dominant coolant of the neutral ISM in galaxies and primarily arises from photodissociation regions (PDRs, e.g., \citealt{1996ApJ...465..738I}; \citealt{1991ApJ...373..423S}). \citet{2017ApJ...834L..16U} found that the [C\,\textsc{ii}] emission is relatively strong compared to the infrared luminosity and [N\,\textsc{ii}] emission and suggested that these characteristics of the ISM are influenced by the location within the giant LAB.
These previous works have gradually uncovered the hidden aspects of LAB1 including the dust-obscured star formation and the nature of the ISM. However, the sensitivity and resolution of the observations were limited.

Here we present results from newly obtained deep [C\,\textsc{ii}], CO(4--3) and dust continuum observations of LAB1 in conjunction with Ly\,$\alpha$ observations. In Section~\ref{sec:obs} we detail the observations and data reduction. In Section~\ref{sec:dust} we describe the source decomposition and flux measurements of the dust continuum emission. In Section~\ref{sec:cii} we derive various properties of the [C\,\textsc{ii}] and CO(4--3) emission, including kinematics and counterpart identification. We discuss the ISM nature, the phase of galaxy assembly, and powering sources of the Ly\,$\alpha$ emission in Section~\ref{sec:discussion}, and present our conclusions in Section~~\ref{sec:conclusion}.
Throughout the paper, we adopt a cosmology with 
$\Omega_{\rm m}=0.3, \Omega_\Lambda=0.7$, and H$_0$=70 km s$^{-1}$ Mpc$^{-1}$.

\section{Observations and Data Reduction} \label{sec:obs}

\subsection{Overview of ALMA Data}

\begin{figure*}
\epsscale{0.77}
\plotone{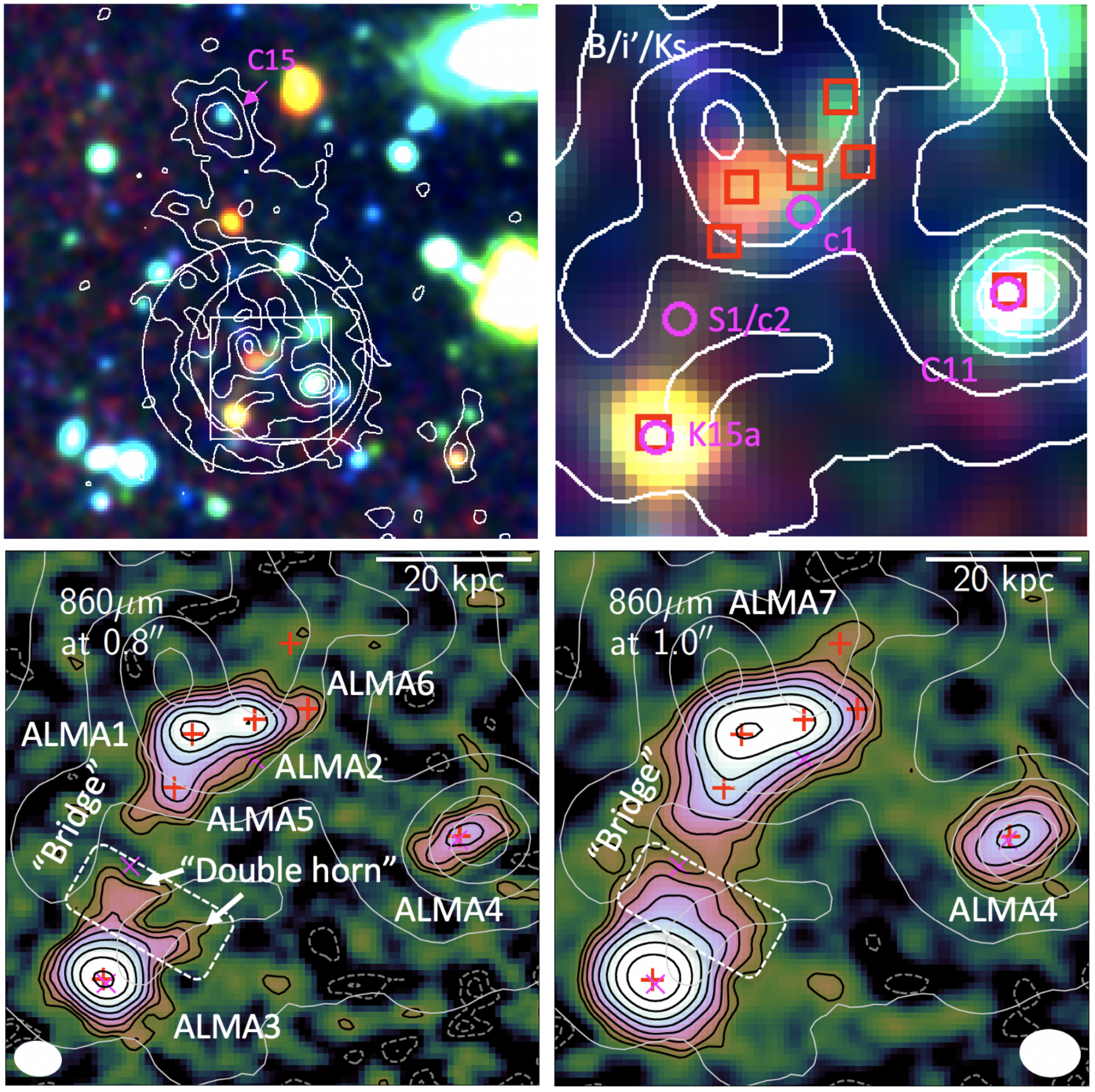}
\caption{
Top left panel shows a false color map of LAB1 (Blue--Subaru/Suprime-Cam, {\it B}-band; Green-- Subaru/Suprime-Cam, {\it i$^{\prime}$}-band; Red--Subaru/MOIRCS, {\it K$_{\rm s}$}-band, \citealt{2004AJ....128.2073H}). White contours show Ly\,$\alpha$ surface brightness of $\mu=[5, 14, 26, 40, 56, 73] \times 10^{-19}$\,erg\,s$^{-1}$\,cm$^{-2}$\,arcsec$^{-2}$~\AA$^{-1}$ averaged over 4969--5004\,\AA. ALMA fields of view are also shown by the two circles (small: Band7, 8, large: Band3).
A zoomed $9^{\prime\prime}\times9^{\prime\prime}$ region (white box in the top left) is shown in a top right panel. 
Bottom two panels show 860\,$\mu$m continuum images of the same field. Thick contours show [$\pm$1.5$^2$, $\pm$1.5$^3$, ...] $\times \sigma_{\rm center}$ where $\sigma_{\rm center}$ is the rms level at the phase center in each map.
ALMA sources (red boxes or crosses) and other known $z\approx3.1$ galaxies with [O\,III] 5008 line detections are labeled (magenta circles or crosses).
The 860\,$\mu$m images reveal widely extended dust components.
\label{fig:cont1}
}
\end{figure*}

\begin{deluxetable*}{cccccccc}
\tabletypesize{\scriptsize}
\tablecaption{Continuum properties of ALMA sources in LAB1}
\tablewidth{0pt}
\tablehead{
\colhead{Source} & \colhead{R.A.} & \colhead{Dec.} & \colhead{$S_{\rm 656\mu m}^{0.8^{\prime\prime}}$} & \colhead{$S_{\rm 850\mu m}^{0.8^{\prime\prime}}$} & \colhead{$S_{\rm 850\mu m}^{1.0^{\prime\prime}}$} & \colhead{$S_{\rm 2.82mm}^{1.4^{\prime\prime}}$} & \colhead{Other ID}\\
 \colhead{} & \colhead{[h m s]} & \colhead{[$\circ$ $^{\prime}$ $^{\prime\prime}$]} & \colhead{[mJy]} & \colhead{[mJy]} & \colhead{[mJy]} & \colhead{[mJy]} 
}
\startdata
ALMA1 & 22 17 26.01 & +00 12 36.4 & 1.56 $\pm$ 0.20$^\dagger$ & 0.50 $\pm$ 0.02  & 0.51 $\pm$ 0.01  & $<$ 0.010         & ALMA-a$^{1}$, {\rm K15c}$^{2}$ \\
ALMA2 & 22 17 25.94 & +00 12 36.7 & 0.70 $\pm$ 0.15         & 0.38 $\pm$ 0.02  & 0.48 $\pm$ 0.01$^\dagger$  & ---$^\dagger$               & ALMA-b$^{1}$\\
ALMA3 & 22 17 26.11 & +00 12 32.3 & 2.49 $\pm$ 0.31         & 0.86 $\pm$ 0.02  &  0.97 $\pm$ 0.01  & 0.022 $\pm$ 0.006 & ALMA-c$^{1}$, {\rm K15a}$^{2}$\\
ALMA4 & 22 17 25.71 & +00 12 34.7 & 0.56 $\pm$ 0.16         & 0.20 $\pm$ 0.01  & 0.23 $\pm$ 0.005 & $<$ 0.010         & C11$^{3}$\\
ALMA5 & 22 17 26.03 & +00 12 35.5 & $<$ 0.17 $^\dagger$       & 0.17 $\pm$ 0.01  & 0.17 $\pm$ 0.01  & ---$^\ddagger$               &  \\
ALMA6 & 22 17 25.88 & +00 12 36.9 & $<$ 0.16                & 0.05 $\pm$ 0.005 & 0.03 $\pm$ 0.002$^\dagger$ & ---$^\ddagger$               &  \\
ALMA7 & 22 17 25.90 & +00 12 38.0 & $<$ 0.16                & $<$ 0.04         & 0.06 $\pm$ 0.006 & --- $^\ddagger$            &  \\
Bridge & --- & --- & --- & --- & 0.20 $\pm$ 0.05 & --- & --- \\
\enddata
\tablecomments{
References are 1: \citet{2016ApJ...832...37G}, 2: \citet{2016MNRAS.455.3333K}, 3: \citet{1998ApJ...492..428S}. 
$\dagger$ Some fraction of fluxes of fainter sources may be underestimated due to difficulty of source deblending.
$\ddagger$ Upper limits are shown only for the brightest three sources since the beam size is too large to isolate limits on fainter sources.
}
\label{table:cont}
\end{deluxetable*}

%
%

LAB1 has been targeted by several ALMA projects. We utilized observations in ALMA Band\,3, Band\,7, and Band\,8, combining both newly obtained data and archival data.
The top left panel of Figure~\ref{fig:cont1} shows a false color map taken with Subaru/Suprime-Cam (\citealt{2004AJ....128.2073H}) and MOIRCS (\citealt{2012ApJ...750..116U}) with Ly\,$\alpha$ contours observed by the Multi Unit Spectroscopic Explorer (MUSE; \citealt{2010SPIE.7735E..08B}). Two LABs were individually identified and labelled as LAB1 and LAB8, originally (\citealt{2000ApJ...532..170S}; \citealt{2004AJ....128..569M}). A deeper Ly\,$\alpha$ map has uncovered that the two LABs are connected to each other (\citealt{2016ApJ...832...37G}; \citealt{2020A&A...642A..55H}; \citealt{2021MNRAS.502.2389L}).
As shown in Fig.~\ref{fig:cont1}, observations in ALMA Band\,7 and Band\,8 cover LAB1 almost entirely, while ALMA Band\,3 observations cover both LAB1 and LAB8 within the field of view. In this paper, we focus on LAB1. The details of the ALMA observations and data reductions are as follows.

\subsection{ALMA Band\,8}

The first observations of LAB1 in ALMA Band\,8 were carried out in Cycle\,2 as reported in \citet{2017ApJ...834L..16U} (ID: 2013.1.00159.S; PI: H. Umehata). In the pilot survey, the central coordinate was ($\alpha$, $\delta$) = ($22^{\rm h}17^{\rm m}26.0^{\rm s}$, $0^{\rm h}12^{\rm m}37.5^{\rm s}$) (ICRS) and on-source time was 4.5~min. Subsequently we performed deeper imaging in ALMA Cycle\,5 (Program ID: 2017.1.01209.S, PI. H. Umehata). We set central coordinates of ($\alpha$, $\delta$) = ($22^{\rm h}17^{\rm m}25.9^{\rm s}$, $0^{\rm h}12^{\rm m}36.3^{\rm s}$) (ICRS). We note that the pointing was slightly shifted, considering the results of the dust-continuum at Band~7 (\citealt{2016ApJ...832...37G}; \citealt{2017ApJ...850..178A}) and the results of the pilot [C\,\textsc{ii}] observations, which were available at the time of the preparation. 

The observations were carried out between May and July in 2018 with 44--45 available 12\,m antennas in the C43--1 and C43--2 configurations. The baseline lengths span from 15\,m to 313\,m. The precipitable water vapor (PWV) was in the range 0.2--0.6\,mm and the weather conditions were excellent or acceptable for Band~8 observations. The exposure time totalled 116.5\,min after combining Cycle\,2 and Cycle\,5 observations. We used the FDM correlator and set the central frequencies of four spectral windows as 451.51, 453.09, 463.14, and 464.77\,GHz. Each spectral window had 1920 channels and the resultant channel width was approximately 1\,MHz. The quasars J2253+1608 or J2258$-$2758 were observed for bandpass and flux calibration and the quasar J2226+0052 was utilized for phase calibration.

Each measurement set was calibrated using the Common
Astronomy Software Application ({\sc casa}) v\,5.1.1 (\citealt{2007ASPC..376..127M}), utilizing the standard reduction pipeline. Imaging the $uv$--data was performed using {\sc casa} v\,5.6.1. 
We first Fourier transformed the $uv$--data to obtain a dirty cube using the {\sc tclean} task, adopting natural weighting. 
We then analyzed the cube to extract the [C\,\textsc{ii}] emission. For this, we focused  on two of the four spectral windows which covers a frequency range of 462.16 to 465.62\,GHz contiguously, corresponding to frequencies of [C\,\textsc{ii}] at $z\simeq3.1$.
We measured the rms level for each channel in a line-free region and cleaned to 2$\sigma$, placing masks for the [C\,\textsc{ii}]-emitting regions. Imaging with natural weighting yields a typical synthesized beam of $0.75^{\prime\prime}\times0.63^{\prime\prime}$ (P.A.=$-80^\circ$). This cube is called the $0.8^{\prime\prime}$ cube in this paper. We applied the {\sc imcontsub} task to subtract continuum emission in the image plane. The resultant rms level at the phase center is 0.50\,mJy\,beam$^{-1}$ with a 20\,km\,s$^{-1}$ velocity bin, while some frequency ranges have relatively higher rms levels, affected by lower atmospheric transmission in Band\,8. The primary beam response {\bf was corrected}. 
To extract spatially extended emission and also perform an angular-resolution-matched comparison with CO(4--3) and 860\,$\mu$m data, we also produce $1.0^{\prime\prime}$ and $1.4^{\prime\prime}$ cubes in the same way but applying varying $uv$-tapering and 80\,km\,s$^{-1}$ velocity bins. The two cubes have synthesized beams and typical rms levels of $0.97^{\prime\prime}\times0.84^{\prime\prime}$ (P.A.=$-83^\circ$) and 0.30\,mJy\,beam$^{-1}$, and $1.44^{\prime\prime}\times1.32^{\prime\prime}$ (P.A.=$-87^\circ$) and 0.39\,mJy\,beam$^{-1}$, respectively.

A continuum image was also created in the same way, using line-free channels in all spectral windows. The map has a representative frequency 458.097\,GHz and the synthesized beam of $0.75^{\prime\prime}\times0.63^{\prime\prime}$ (P.A.=$-81^\circ$). The 1$\sigma$ sensitivity at the phase center is 52\,$\mu$Jy beam$^{-1}$. 

\subsection{ALMA Band\,7}

Following the first two observations in ALMA Cycle\,2 (Program ID: 2013.1.00704.S, PI: Y. Matsuda, Program ID: 2013.1.00922.S, \citealt{2016ApJ...832...37G}; \citealt{2017ApJ...850..178A}), LAB1 was further observed in ALMA Band~7 by two projects.
In ALMA Cycle\,4 (Program ID: 2016.1.01134.S, PI. J. Geach), observations were performed to obtain a deeper continuum map at 850\,$\mu$m, centered at ($\alpha$, $\delta$) = ($22^{\rm h}17^{\rm m}26.0^{\rm s}$, $0^{\rm h}12^{\rm m}34.7^{\rm s}$) (ICRS). The observations were carried out on 4 and 5 April  2017, using 38--39 usable 12\,m antennas under good weather conditions. The C40--1 array configuration utilized was the most compact configuration at the time (baseline lengths of 15 to 279\,m), which was suitable to detect extended components. The representative frequency was 354.60~GHz and the total on-source time was 82~min. The quasars, J2148+0657 and J2232+1143 were observed for calibration. Each measurement set was calibrated in {\sc casa} v\,4.7.2, utilizing the standard reduction pipeline.

LAB1 was also observed in ALMA Band~7 as a part of the Cycle\,5 project (Program ID: 2017.1.01209.S, PI: H. Umehata) to detect dust continuum and [N\,\textsc{ii}]\,205~$\mu$m emission. We will report the result of [N\,\textsc{ii}]\,205~$\mu$m emission in a separate paper (H. Umehata et al. in preparation). Observations were carried out on 30 and 31 May 2018, using 45--46 available 12\,m antennas, under good weather conditions. The C43-2 array configuration was used, which resulted in a range of baselines of 15 to 314\,m. The central position was ($\alpha$, $\delta$) = ($22^{\rm h}17^{\rm m}26.0^{\rm s}$, $0^{\rm h}12^{\rm m}37.1^{\rm s}$) (ICRS). The total on-source time was 2.4\,hours, which was divided into three execution blocks. The TDM correlator was utilized to have a representative frequency of 356.3735\,GHz. The two sidebands have a frequency gap of 8\,GHz between them, and each sideband has two spectral windows with 1.875\,GHz bandwidth and 31.25\,MHz resolution. The nearby quasar J2226+0052 was observed regularly to calibrate amplitude and phase, while J2253+1608 was observed for bandpass, pointing, and absolute flux calibration. Each measurement set was calibrated in {\sc casa} v\,5.1.1, utilizing the standard reduction pipeline.

All available Band~7 data were mapped using the {\sc tclean} task in {\sc casa}. As for Band\,8, maps with three angular resolutions were created. We excluded channels which cover the redshifted [N~\textsc{ii}]\,205\,$\mu$m emission in the continuum imaging. We then cleaned to 2$\sigma$, masking bright sources. The resulting image has a representative frequency of 347.565\,GHz.
Imaging with the Briggs parameter 0.5 resulted in the synthesized beam $0.79^{\prime\prime}\times0.59^{\prime\prime}$ (P.A.=$83^\circ$) and we refer to it the ``$0.8^{\prime\prime}$ image''. Images obtained using natural weighting provide images with the synthesized beam $1.02^{\prime\prime}\times0.80^{\prime\prime}$ (P.A.=$-81^\circ$) and $1.54^{\prime\prime}\times1.34^{\prime\prime}$ (P.A.=$-81^\circ$) without and with tapering, respectively. We refer to these as the $1.0^{\prime\prime}$ and $1.4^{\prime\prime}$ images. The 1$\sigma$ sensitivity is 11, 12, and 16\,$\mu$Jy beam$^{-1}$ at the phase center, for the $0.8^{\prime\prime}$, $1.0^{\prime\prime}$ and $1.4^{\prime\prime}$ images, respectively.

\subsection{ALMA Band\,3}

LAB1 was observed in ALMA Band\,3 in its Cycle\,4 (Program ID: 2016.1.00485.S, PI. N. Hine). The central position was ($\alpha$, $\delta$) = ($22^{\rm h}17^{\rm m}26.0^{\rm s}$, $0^{\rm h}12^{\rm m}37.6^{\rm s}$) (ICRS). Observations were carried out in December 2016 with 41--46 usable 12\,m antennas. The baseline lengths ranged from 15 to 243\,m. The total on source time was 4.6~hours, divided into 6 individual EBs. The quasars J2148+0657 and J2226+0052 were observed for pointing, amplitude, bandpass, and phase calibration. The absolute flux scale was set using observation of Neptune. Each measurement set was calibrated in {\sc casa} v\,4.7.0, utilizing the standard reduction pipeline. Data were mapped using the {\sc tclean} task in {\sc casa} with natural weighting. Our primary target was the CO(4-3) line at $z\sim3.1$, and a cube was created and cleaned to $2\sigma$ with bright sources masked. The resultant size of the synthesized beam is $1.53^{\prime\prime}\times1.30^{\prime\prime}$ (P.A.=$-33^\circ$) at 112.52\,GHz. The typical rms level at the phase center is 0.70\,$\mu$Jy\,beam$^{-1}$ with 80\,km\,s$^{-1}$ velocity bins. We also made a continuum map from line-free channels using {\sc tclean}. The map has a synthesized beam of $1.61^{\prime\prime}\times1.39^{\prime\prime}$ (P.A.=$-33^\circ$) and $1\sigma$ sensitivity of 4.9\,$\mu$Jy at 106.28\,GHz.

\subsection{MUSE}

LAB1 has been observed by MUSE on UT4 of the Very Large Telescope in three programs (094.A-0605 PI: M. Hayes, 095.A-0570 PI: R. Bower, 097.A-0831 PI: M. Hayes), which provides a three-dimensional data cube containing Ly\,$\alpha$ emission (\citealt{2016ApJ...832...37G}; \citealt{2020A&A...642A..55H}). The typical individual exposure times are 1500\,sec and the total on-source time is 17.6\,hours. Seeing was typically about 1~arcsec (see \citealt{2020A&A...642A..55H} for details). 
Data were reduced with the MUSE pipeline (\citealt{2016ascl.soft10004W}), following standard procedures. Flat fielding and sky subtraction were performed with an additional correction to homogenize the illumination across the field (e.g., \citealt{2017MNRAS.467.3140S}; \citealt{2019Sci...366...97U}).
The resultant cube has wavelength bins of 1.25 \AA.
In this work, all wavelengths were finally specified in vacuum with {\sc airtovac} in the {\sc mpdaf} package (\citealt{2016ascl.soft11003B}; \citealt{2017arXiv171003554P}) using the relation of \citet{1996ApOpt..35.1566C}. We use the vacuum wavelength 1215.67\,\AA~for Ly\,$\alpha$.

\section {Dust continuum in LAB1} \label{sec:dust}

Continuum maps of LAB1 at 860\,$\mu$m are shown in Figure~\ref{fig:cont1}, compared with the Ly\,$\alpha$ emission. As shown, the newly obtained ALMA continuum maps in Band\,7 have changed our view of this system; multiple cold dust emission components are discovered across LAB1, including three galaxies previously identified by \citet{2016ApJ...832...37G},

To isolate the emission from each component, bright sources were sequentially modeled and subtracted using {\sc casa}/{\sc imfit}. This results in identification of seven dusty galaxies, ALMA1 through ALMA7 as labeled in Figure~\ref{fig:chan1} (see Figure~\ref{fig:decomp} for model and residual images). \citet{2016ApJ...832...37G} suggested the existence of a ``tail'' connected to ALMA1, lying to the south. This may be associated with ALMA5, although the low signal-to-noise ratio of the image presented in \citet{2016ApJ...832...37G} precludes a definitive conclusion.
There is an additional dusty component located between ALMA3 and ALMA5. This ``Bridge'' region shows a double horn shape (Figure~\ref{fig:cont1}).

The 860\,$\mu$m flux densities of the seven components were measured using {\sc casa/imfit} using the 0.8$^{\prime\prime}$ and 1.0$^{\prime\prime}$ images as summarized in Table~\ref{table:cont}. Following \citet{2016ApJ...832...37G}, we also measured the total flux density of the 860\,$\mu$m signal above the 3$\sigma$ level for the complex including six galaxies (except for ALMA4) and the Bridge (Table~\ref{table:cont}). The sum of the individual galaxies and bridge is consistent with the measurement for the whole complex, which suggests that the eight components account for most of the 860\,$\mu$m flux density.
%
Band\,8 and Band\,3 photometry are also summarized in Table~\ref{table:cont}.
Two [O\,\textsc{iii}] emitters at $z\approx3.1$, c1 and c2 (\citealt{2016ApJ...832...37G}; \citealt{2021MNRAS.502.2389L}) are not individually detected in dust continuum.

The integrated flux of all components is $S_{860}= 2.66 \pm 0.11$~mJy in the 1.4$^{\prime\prime}$ map (Table~\ref{table:ism}). This value is $\sim$40\% larger than the previously reported flux, $S_{\rm 860}=1.86\pm 0.06$\,mJy (\citealt{2016ApJ...832...37G}). Newly identified relatively extended and/or faint components account for the increase.
The updated ALMA-based flux is found to fall in the range of deboosted SCUBA2 measurements ($S_{850}=3.6\pm1.2$~mJy, \citealt{2017MNRAS.465.1789G}).
We note that it remains unclear whether or not all of the emission observed by the single-dish telescope is completely recovered by ALMA or not, as the remaining uncertainty ($0.7\pm1.2$~mJy) indicates.

\section{[CII] and CO emissions in LAB1} \label{sec:cii}

\begin{figure*}
\epsscale{0.95}
\plotone{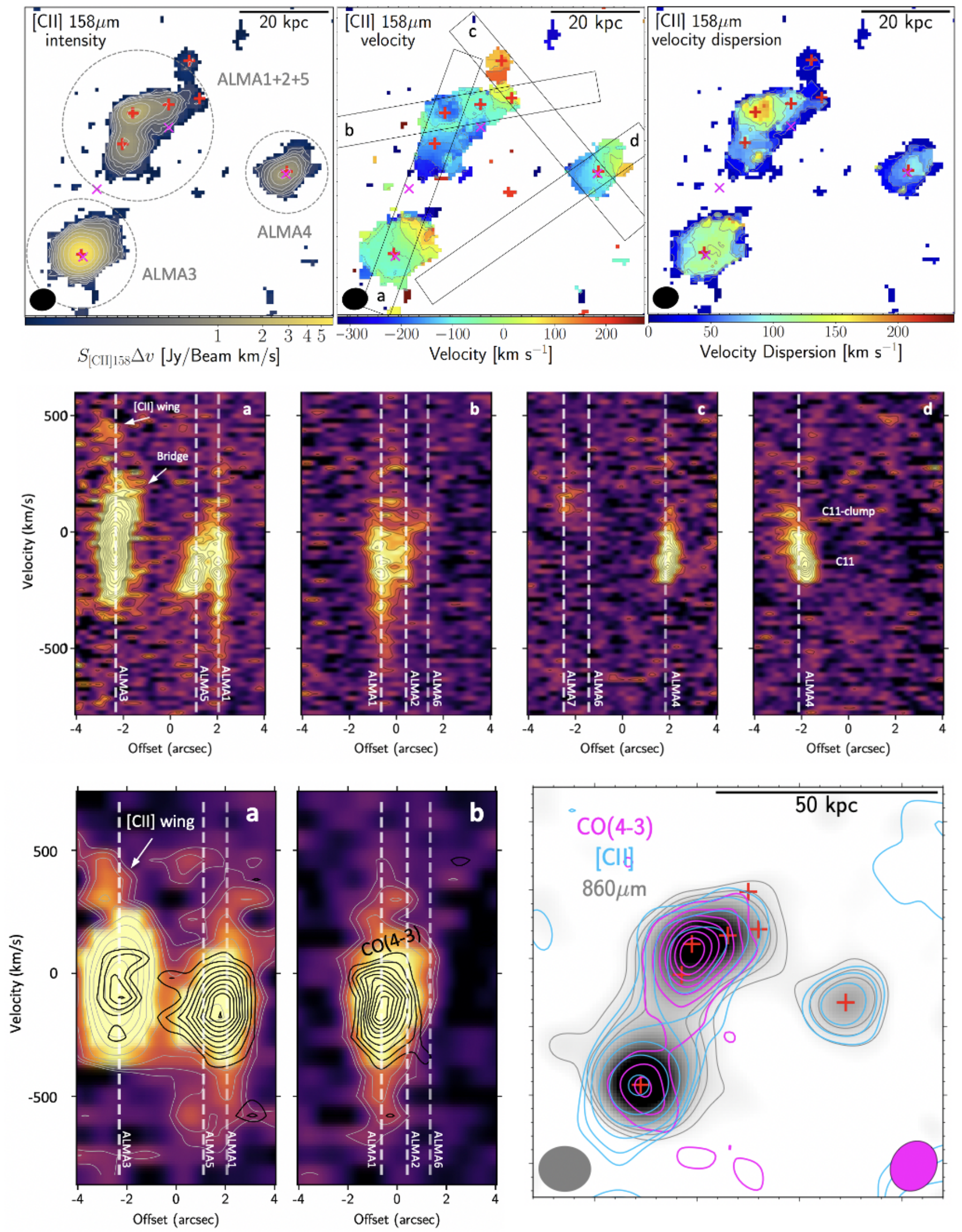}
\caption{
Top row panels show moment maps generated using a {\it rendered} cube  (see Section~\ref{sec:overview} for more details) are displayed to show overall trends of  [C\,\textsc{ii}] emission in LAB1. The positions of dust continuum sources and [O\,\textsc{iii}] emitters are shown.
Contours are $S_{\rm [CII]158}\Delta v=$  0.1$\times$[1.5, 1.5$^2$, ..., 1.5$^{11}$   ]\,Jy\,beam$^{-1}$\,km\,s$^{-1}$, in steps of 50\,km\,s$^{-1}$ and 25\,km\,s$^{-1}$ relative to $z=3.100$, from left to right, respectively. 
Middle row panels show [C\,\textsc{ii}] P--V diagrams at 0.8$^{\prime\prime}$ along four slits ($a\sim d$). Contours are in steps of $1\sigma_{\rm center}$ starting at $2\sigma_{\rm center}$. 
ALMA4 and ALMA5 show velocity gradient nearly along the slit, which suggests that they have a rotating gas-disk. ALMA4 also has a clump-like structure (``C11-clump''). Multiple-merging events are ongoing among ALMA1, ALMA2, ALMA4, and ALMA6. 
ALMA3 is accompanied by relatively faint and extended emissions, which is the most remarkable at higher velocities (``[C\,\textsc{ii}] wings'').
Bottom left two panels show [C\,\textsc{ii}] P--V diagrams at 1.4$^{\prime\prime}$. CO(4-3) emissions are superposed. Contours show [1.5$^3$, 1.5$^2$, ...] $\times \sigma_{\rm center}$ and [2, 3, ...] $\times \sigma_{\rm center}$, respectively. A bottom right panel show CO(4-3) and [C\,\textsc{ii}] emission superposed on the 860~$\mu$m map, integrated over the range of $-379$\,km\,s$^{-1}$ to 101\,km\,s$^{-1}$.
CO(4-3) and [C~\textsc{ii}] contours are [2, 4, 8, 16, 32, 64]$\times \sigma_{\rm center}$, while 860~$\mu$m contours stand for [2, 3, 4, 5]$\times \sigma_{\rm center}$. 
The kinematics traced by CO(4-3) is similar to that of [C\,\textsc{ii}]. There is no detectable CO(4-3) emission around ALMA4.
}
\label{fig:momentpv}
\end{figure*}

\begin{figure}
\epsscale{1.15}
\plotone{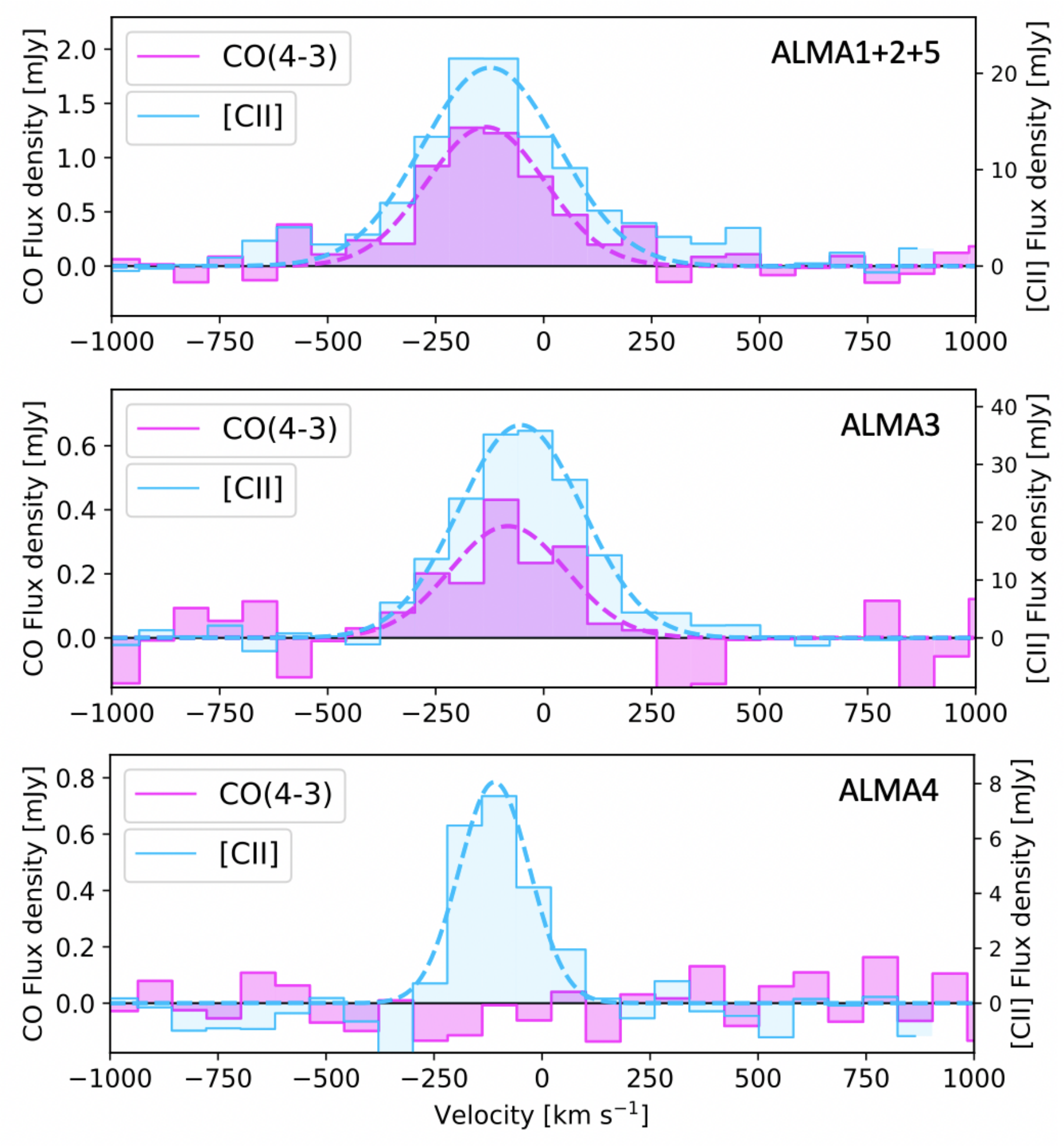}
\caption{
CO(4--3) and [C\,\textsc{ii}] spectra of four regions in LAB1. Velocities are relative to $z=3.100$. Dashed lines show a Gaussian profile fitted to these emissions.
Simultaneous detections of the two emission lines are identified in ALMA1+2+5 and ALMA3, while no CO(4-3) line is detected for ALMA4. 
}
\label{fig:spec}
\end{figure}

\subsection{Overview} \label{sec:overview}

As shown in Figure~\ref{fig:momentpv}, the [C\,\textsc{ii}] emission is widely distributed across LAB1 over an area of $d\simeq100$\,kpc, highlighting complex  morphology in three-dimensional space. All seven components identified in the dust continuum are associated with corresponding [C\,\textsc{ii}] emission, confirming that the dust continuum detected within LAB1 is also at $z\approx3.10$.
The picture of the [C\,\textsc{ii}] line has been significantly updated compared to our pilot survey. \citet{2017ApJ...834L..16U} detected the line only for ALMA3. The deeper observations now confirm that the previous observations also covered the frequency range containing [C\,\textsc{ii}] emission, even though the lower sensitivity prevented actual detection.  

To illustrate the overall trend of the [C\,\textsc{ii}] emission projected to a two-dimensional space, we also created a suite of moment maps. For this purpose, we made a {\it rendered} cube using the naturally weighted, 0.8$^{\prime\prime}$ cube. 
First, voxels that had emission above the $2\sigma$ level, measured for each channel before primary beam correction, was extracted from the cube. Second, among them, if a voxel connects to another extracted voxel in velocity space, the voxel was considered to have emission and left in the box. If not, the voxel was masked. Finally, a primary beam correction was applied for each channel. This process allows us to effectively extract extended emission from the data cube, suppressing the influence of noise. 

Using the rendered cube, the integrated emission map, the flux-weighted velocity map, and the flux-weighted velocity dispersion map were calculated using {\sc casa}/{\sc immoments} task.
%
As shown in Figure~\ref{fig:momentpv}, the extracted [C\,\textsc{ii}] emission is 
generally cospatial with dust continuum components.
The most dominant five sources (ALMA1--ALMA5) are composed of three groups (ALMA1+2+5, ALMA3, and ALMA4). 
The remaining two sources are associated with ALMA1+2+5.
%
The velocity map suggests that the emission has complicated velocity structure in a relatively narrow range of flux-weighted velocity (about $-200$\,km\,s$^{-1}$ to 100\,km\,s$^{-1}$), while some parts of them, such as ALMA4, show a coherent velocity structure. The velocity dispersion map suggests a variation of (apparent) velocity dispersion ($ 50$\,km\,s$^{-1}$ to 200\,km\,s$^{-1}$) among the components, which implies various dynamical states among the cool gas components in LAB1.

The bottom left panel of Figure~\ref{fig:momentpv} shows the spatial distributions of CO(4--3), [C\,\textsc{ii}], and 860\,$\mu$m at 1.4$^{\prime\prime}$ resolution. Spectra are shown in Figure~\ref{fig:spec}. The CO(4--3) emission line is identified in ALMA1+2+5 and ALMA3, while ALMA4 has no detected emission. While their spatial distributions are similar on large scales, they are not identical on smaller scales. There is also a wide variety in line ratios, which implies that there are different ISM states among the galaxies in LAB1. Further line diagnostics are beyond the scope of this paper, but we will present them in an upcoming paper (H. Umehata et al. in preparation).

\subsection{Kinematics of the [CII] and CO emissions} \label{subsec:ciiprop}


Position--velocity (P--V) diagrams at 0.8$^{\prime\prime}$ and 1.4$^{\prime\prime}$ resolution are displayed in Figure~\ref{fig:momentpv}. Each pseudo slit has a width of 0.9$^{\prime\prime}$ to cover the majority of emission along the direction of the velocity gradient, which enables us to discern in more detail the velocity structure and interaction among sources in this complicated system in more detail. 

ALMA1+2+5 encompasses ALMA1, ALMA2, ALMA5, as well as ALMA6, and ALMA7.
As demonstrated, ALMA1 and ALMA2 are located closely each other (with a projected angular separation $1^{\prime\prime}$) and overlap in velocity space. This is also the case for ALMA1 and ALMA5. In the $0.8^{\prime\prime}$ map,
ALMA1 appears to be simultaneously interacting with ALMA2 and ALMA5. 
The moment map shows that velocity dispersion peaks in the region between ALMA1 and ALMA2, which is also suggestive of turbulent nature there. 
ALMA6 and ALMA7 show coherent velocity structure, smoothly connecting with ALMA2, which could be additional merging galaxies or tidal tails associated with ALMA2.
%
In Figure~\ref{fig:momentpv}, slit-a shows a nearly monotonic velocity gradient along the slit at the position of ALMA5 and thus an ordered rotation is indicated.
The velocity structure of the CO(4--3) emission is generally consistent with that of [C\,\textsc{ii}]. The emission peak is the closest to ALMA1, while the CO(4-3) profile is also elongated toward ALMA2 and ALMA5. 

ALMA3 is the brightest galaxy in [C\,\textsc{ii}] in LAB1, and is also covered by slit-a. There is no significant shift in the centroid position between velocity channels spanning about 400\,km\,s$^{-1}$ around the peak. The Bridge corresponds to a protrusion in a redder part, which is likely to account for the apparent velocity gradient evident in the velocity map. The velocity dispersion of ALMA3 is higher than that of other ALMA sources on galaxy scales (Figure~\ref{fig:momentpv}), which suggests that turbulence dominates the kinematics in ALMA3. CO(4--3) emission is also identified in ALMA3; it does not show a velocity shift as for [C\,\textsc{ii}]. 
Slit-a also shows that ALMA3 is accompanied by faint emission with a complicated morphology, including a red ``[C\,\textsc{ii}] wing''. Outflows, tidal tails, or merging satellites could account for the emission. 
%

In the case of ALMA4, there is a nearly monotonic velocity gradient, suggesting a rotating disk. There is a high-velocity clump (``C11-clump''). Since the C11-clump follows the velocity shift, the system would be kinematically dominated by rotation of C11. However, the velocity dispersion peaks between the dust/stellar peak and the clump, and interaction between the C11 and C11-clump is also implied. ALMA4 has no detectable CO(4-3) emission.

\subsection{Radial Profile}

\begin{figure}
\epsscale{1.15}
\plotone{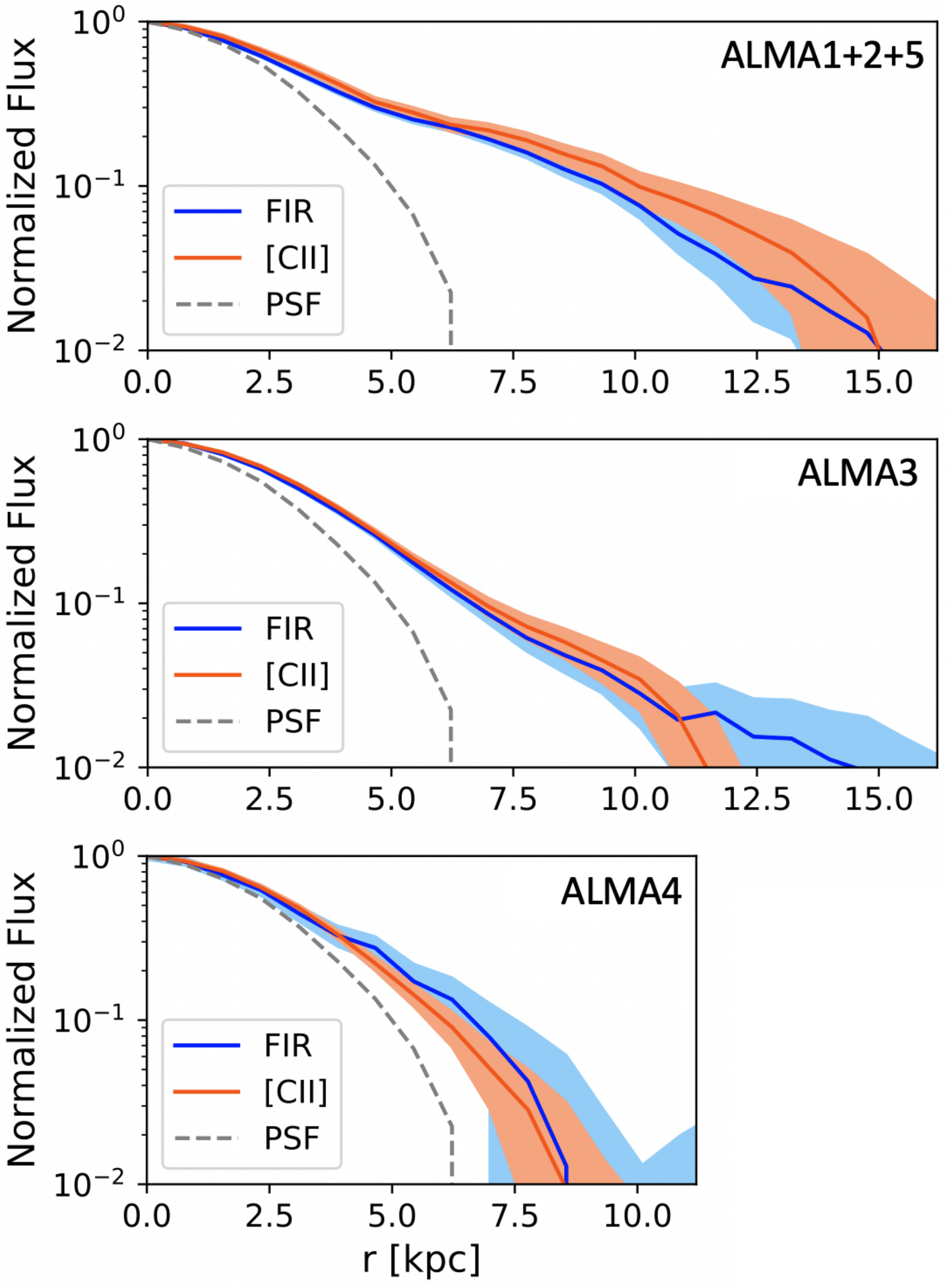}
\caption{
Radial profiles centered at the three bright dusty starbursts in 0.8$^{\prime\prime}$ resolution maps. The 860\,$\mu$m map is used to show the FIR profile. Shaded regions show the photometric uncertainties (not including the calibration errors).
Radial profiles are generally similar between FIR and [C\,\textsc{ii}] up to about 15\,kpc. 
}
\label{fig:radial}
\end{figure}

Figure~\ref{fig:radial} shows radial profiles of the dust and gas components at 0.8$^{\prime\prime}$ resolution traced by the 860\,$\mu$m continuum and [C\,\textsc{ii}] emission, respectively. The [C\,\textsc{ii}] emission lines are integrated over a velocity range optimized for each region (see also Figure~\ref{fig:hst}).
A combined profile for ALMA1+2+5 is measured, centred at the position of ALMA1.
In the case of ALMA3, the Bridge region also contributes in addition to ALMA3 itself. For all cases, both [C\,\textsc{ii}] and dust continuum emission shows similar radial profiles (extending to about 15\,kpc).
The profiles demonstrate that the gas and dust coexist on a large scale in LAB1.
We will discuss the nature of the widely distributed ISM in Section~\ref{sec:discussion}.

\subsection{Counterparts of ALMA-identified populations}

\begin{figure}
\epsscale{1.15}
\plotone{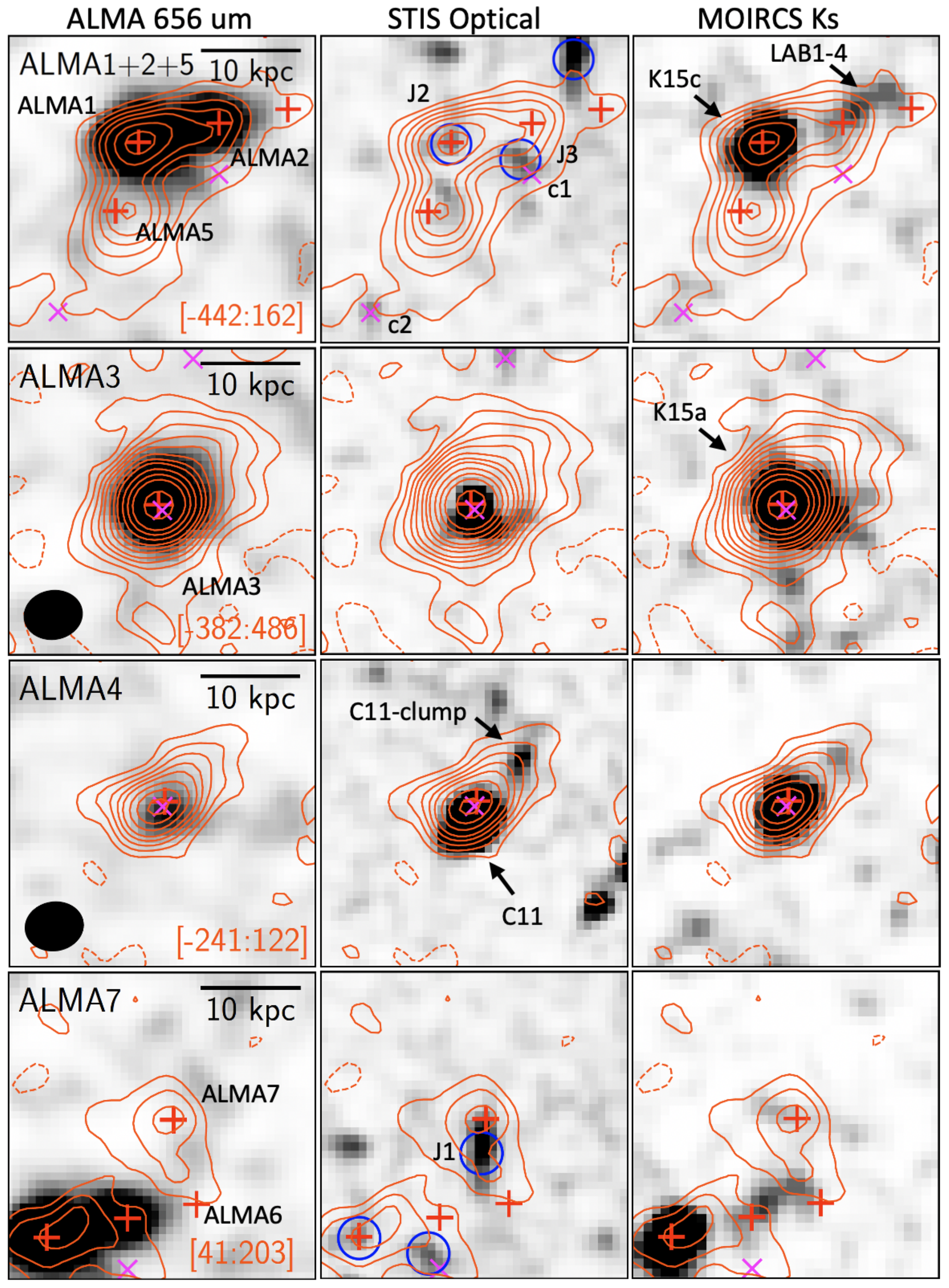}
\caption{
Multi-wavelength images of the four [C~\textsc{ii}]-emitting regions taken with ALMA, {\it HST}/STIS Optical (\citealt{2004ApJ...606...85C}), and Subaru MOIRCS (\citealt{2012ApJ...750..116U}). Orange contours show [-3, 2, 4, 6, 8, 10, 13, 16, 19, 24, 29, 34] $\times\sigma_{\rm center}$ of the integrated [C~\textsc{ii}] emission. The velocity ranges used are shown in each ALMA panel in the unit of [km\,s$^{-1}$]. Crosses show positions of dust continuum (red) and [O~\textsc{iii}] emitters (magenta), while blue circles show faint optical sources identified in \citet{2004ApJ...606...85C} (J1 also has a VLA counterpart as in \citet{2004ApJ...606...85C}). 
There is a variety of appearance in the optical and {\it K$_{\rm s}$} band images among the ALMA-identified components. ALMA5 has no counterpart and highlights the importance of ALMA observations.
}
\label{fig:hst}
\end{figure}

Observations at optical-to-near-infrared wavelengths provide information on stellar components, tracing rest-frame UV-to-optical emission for galaxies at $z\approx3.1$. Figure~\ref{fig:hst} shows an optical image taken with {\it HST} (the Space Telescope Imaging Spectrograph, STIS)\footnote{The image has a pivot wavelength of 5733~\AA.} (\citealt{2004ApJ...606...85C}), a $K_{\rm s}$-band image obtained with the Subaru Multi-Object InfraRed Camera and Spectrograph (MOIRCS) (\citealt{2012ApJ...750..116U}), and dust continuum observations taken in ALMA Band\,8. The [C\,\textsc{ii}] emission is integrated over a velocity range noted in the figure, and known galaxies are labeled (\citealt{2004ApJ...606...85C}; \citealt{2012ApJ...750..116U}; \citealt{2015ApJ...799...38K}; \citealt{2016MNRAS.455.3333K}; \citealt{2016ApJ...832...37G}; \citealt{2021MNRAS.502.2389L}).

ALMA1, ALMA3, and ALMA4 have significant {\it K$_{\rm s}$}-band counterparts, which is suggestive of modestly obscured star-forming galaxies. 
ALMA2, ALMA6, and ALMA7 have possible counterparts (LAB1-4 for ALMA2 and ALMA6 and J1 for ALMA7) with slight offsets\footnote{Positional uncertainties ($\sigma_{\rm pos}$) related to the signal-to-noise ratio (S/N) and the synthesized beam size ($\theta_{\rm beam}$) is described as $\sigma_{\rm pos}\approx \theta_{\rm beam}/(2\times S/N)$ (\citealt{1997PASP..109..166C}). In the case of $\theta_{\rm beam}=0.8^{\prime\prime}$ and S/N=5, $\sigma_{\rm pos}$ is thus $\approx0.08^{\prime\prime}$, which is smaller than the measured offsets.},
 $0.3-0.6^{\prime\prime}$. This might be explained by different degrees of dust extinction in a galaxy, a companion, or a tidal component.
ALMA5 is blank in the rest-frame UV and optical images, in contrast with its bright [C\,\textsc{ii}] emission. This demonstrates the utility of ALMA to identify populations that would otherwise be missed.

\begin{deluxetable*}{llccc}
\tabletypesize{\scriptsize}
\tablecaption{
ISM properties of components in LAB1 at 1.4$^{\prime\prime}$ resolution
}
\tablewidth{0pt}
\tablehead{
\colhead{} & \colhead{}  & \colhead{ALMA1$+$ALMA2$+$ALMA5} & \colhead{ALMA3} & \colhead{ALMA4}
}
\startdata
CO(4-3) & Coordinates (ICRS) & 22:17:26.00 $+$00:12:36.1 & 22:17:26.10 $+$00:12:32.2 & --- \\ 
 & $z_{\rm CO(4-3)}$ & 3.0982 $\pm$ 0.0002 & 3.0989 $\pm$ 0.0004 & --- \\
 & Diameter $^a$\,[kpc] & (12.7$\pm$1.8)$\times$(7.5$\pm$1.5) & point source  & --- \\ 
 & FWHM\,[km\,s$^{-1}$] & 322 $\pm$ 29 & 316 $\pm$ 65 & --- \\
 & $S\Delta v$\,[Jy\,km\,s$^{-1}$] & 0.44 $\pm$ 0.03 & 0.12 $\pm$ 0.02 & $<$0.03$^b$ \\
 & L$_{\rm CO(4-3)}$\,[10$^{7}$\,L$_\odot$] & 3.6 $\pm$ 0.3 & 1.0 $\pm$ 0.2 & $<$0.3 \\
 & M$_{\rm gas}^{\rm CO(4-3)}$ $^c$\,[10$^{10}$\,M$_\odot$] & 6.8 $\pm$ 1.5 & 1.7 $\pm$ 0.5 & $<$0.5 \\
  &  &  &  &  \\ 
{\rm [C\,\textsc{ii}]} & Coordinates (ICRS) & 22:17:26.00 $+$00:12:36.0 & 22:17:26.11 $+$00:12:32.3  & 22:17:25.71 $+$00:12:34.6 \\  
 & $z_{\rm [CII]}$ & 3.0983 $\pm$ 0.0002 & 3.0993 $\pm$ 0.0001 & 3.0985 $\pm$ 0.0001 \\
 & Diameter $^a$\,[kpc] & $(17.2\pm2.9)\times(7.3\pm2.5)$ & $(8.5\pm1.0)\times(4.6\pm1.5)$  & point source \\   
 & FWHM\,[km\,s$^{-1}$] & 370 $\pm$ 30 & 337 $\pm$ 11 & 193 $\pm$ 24 \\
 & $S\Delta v$\,[Jy\,km\,s$^{-1}$] & 8.12 $\pm$ 0.57 & 13.20 $\pm$ 0.38 & 1.65 $\pm$ 0.18 \\
 & L$_{\rm [CII]}$\,[10$^{8}$\,L$_\odot$] & 27.3 $\pm$ 1.9 & 44.4 $\pm$ 1.3 & 5.6 $\pm$ 0.6 \\   
 &  &  &  &  \\ 
860~$\mu$m & Coordinates (ICRS) & 22:17:25.99 $+$00:12:36.4 & 22:17:26.10 $+$00:12:32.4  & 22:17:25.71 $+$00:12:34.7 \\
 &  Diameter $^a$\,[kpc] & $(15.0\pm1.4)\times(6.5\pm1.2)$ & $(8.9\pm1.3)\times(6.6\pm1.8)$  & $(6.9\pm0.8)\times(3.1\pm2.3)$ \\
 & S$_{850}$\,[mJy] & 1.27$\pm$0.09 & 1.17$\pm$0.07 & 0.22$\pm$0.01 \\
 & M$_{\rm dust}^{850}$ $^d$\,[10$^{8}$\,M$_\odot$] & $\approx$2.0 & $\approx$1.9 & $\approx$0.4 \\
 & L$_{\rm IR[8-1000]}$\,[10$^{12}$\,L$_\odot$] $^e$ & 1.05 $\pm$ 0.34& 0.97$\pm$0.31 & 0.18$\pm$0.06 \\
 & SFR$_{\rm IR}$\,[M$_\odot$\,yr$^{-1}$] $^e$ & 110$\pm$40 & 100$\pm$30 & 20$\pm$6 \\ 
\enddata
\tablecomments{
CO and [C\,\textsc{ii}] spectra are extracted using apertures shown in Figure~\ref{fig:momentpv} (Figure~\ref{fig:spec}).
Properties are generally derived via fits with a single Gaussian profile on the spectra, while {\sc casa/imfit} task is also utilized to derive coordinates and sizes. For the 860\,$\mu$m data, {\sc casa/imfit} are utilized to derive properties.
$^a$ deconvolved major and minor axes of FWHM measured by {\sc casa/imfit} are presented.
$^b$ $3\sigma$ point source limit with an assumption of a velocity width (320\,km\,s$^{-1}$).
$^c$ brightness temperature ratio $r_{41}=0.61\pm0.13$ (\citealt{2020ApJ...902..109B}) and a CO-to-H$_2$ conversion factor $\alpha_{\rm CO}=3.6$ (as in \citet{2020ApJ...902..110D}) are assumed.
$^d$ $M_{\rm d}$ values are calculated assuming modified black-body radiation with $\beta=1.8$ and $T_{\rm d} =40$\,K.
$^e$ L$_{\rm IR}$ and SFR$_{\rm IR}$ are derived in a manner presented in \citet{2018PASJ...70...65U} assuming a Chabrier IMF.
}
\label{table:ism}
\end{deluxetable*}

\subsection{Measurements of properties}\label{sec:mesurement}


Since it is not straightforward to isolate line fluxes from the cube in this crowded region, we derive line properties for groups on the basis of 1.4$^{\prime\prime}$ resolution data.
The line flux, redshift, and the FWHM of the line profile are calculated from a Gaussian fit to the extracted spectra shown in Figure~\ref{fig:spec}. The velocity-integrated maps (Figure~\ref{fig:momentpv}) are used to measure sizes of the emissions using {\sc casa/imfit}. Derived properties are summarized in Table~\ref{table:ism}. 
We also extracted [C\,\textsc{ii}] spectra from the 1.0$^{\prime\prime}$ cube at the positions of seven ALMA sources (ALMA1 through ALMA7) to measure line properties and derive [C\,\textsc{ii}] line for each (Appendix~\ref{sec.A.ciitable}).

%


A suite of other physical properties were also estimated.
IR luminosity ($L_{\rm IR}$) and IR-based SFR (SFR$_{\rm IR}$) were estimated, scaled 860\,$\mu$m fluxes using SED templates of ALESS SMGs (\citealt{2017ApJ...840...78D}) (for more details, see \citet{2018PASJ...70...65U}).
As summarized in Table~\ref{table:ism}, ALMA1+2+5, ALMA3, and ALMA4 have SFR$_{\rm IR}$ of $\sim$110, $\sim$100, $\sim$20\,M$_\odot$\,yr$^{-1}$, respectively. These estimates, however, are based on single point photometry and thus remain uncertain. 

The CO(4--3) line intensities for ALMA1+2+5 and ALMA3 are derived to be $S \Delta {\rm v}=0.44\pm0.03$, $0.12\pm0.02$\,Jy\,km\,s$^{-1}$, respectively. 
We can compare these to earlier attempts to observe the CO(4--3) line from LAB1. \citet{2004ApJ...606...85C} obtained a marginal detection of the line using the Owens Valley Radio Observatory Millimeter Array and reported a line intensity $S {\Delta}{\rm v}<2.5$\,Jy\,km\,s$^{-1}$, while \citet{{2012ApJ...744..178Y}} reported a $3\sigma$ upper limit $S \Delta {\rm v}<0.62$\,Jy\,km\,s$^{-1}$ using the Plateau de Bure Interferometer (PdBI). The ALMA observation, which has a better angular resolution, thus confirms that the CO(4--3) emission had evaded detection due to the insufficient sensitivity of these observations.
The CO(4--3) line luminosities were calculated following \citet{2005ARA&A..43..677S}. To estimate molecular gas mass on the basis of the CO(4-3) line, a brightness temperature ratio $r_{41}=L^{\prime}_{\rm CO(4-3)}/L^{\prime}_{\rm CO(1-0)}$ and a CO-to-H$_2$ conversion factor $\alpha_{\rm CO}$ must be assumed. The derived IR luminosity range, log ($L_{\rm IR}$/L$_\odot$) = 11.3--12.0, is comparable to, or somewhat lower than, that of the ASPECS-LP sample at $z=2.0-4.0$ (log ($L_{\rm IR}$/L$_\odot$) = 11.6--12.9, \citealt{2020ApJ...902..109B}). Hence we adopt the brightness temperature ratio $r_{41}=0.61\pm0.13$, which is derived for the $z>2$ ASPECS galaxies, and $\alpha_{\rm CO}=3.6$\,$M_\odot$ (K\,km\,s$^{-1}$\,pc$^2$)$^{-1}$, which is adopted for the galaxies (\citealt{2020ApJ...902..110D}). Estimated molecular gas masses are $M_{\rm gas}=(6.8\pm1.5$) and ($1.8\pm0.5) \times 10^{10}$\,$M_\odot$ for ALMA1+2+5 and ALMA3, respectively. About 80$\%$ of the molecular gas is concentrated in ALMA1+2+5. Measured values are given in Table~\ref{table:ism}, including an upper limit for ALMA4.

\section{Discussion} \label{sec:discussion}


\subsection{The origin of the [CII] emission} \label{subsec:ciiratio}


\begin{figure*}
\epsscale{1.15}
\plotone{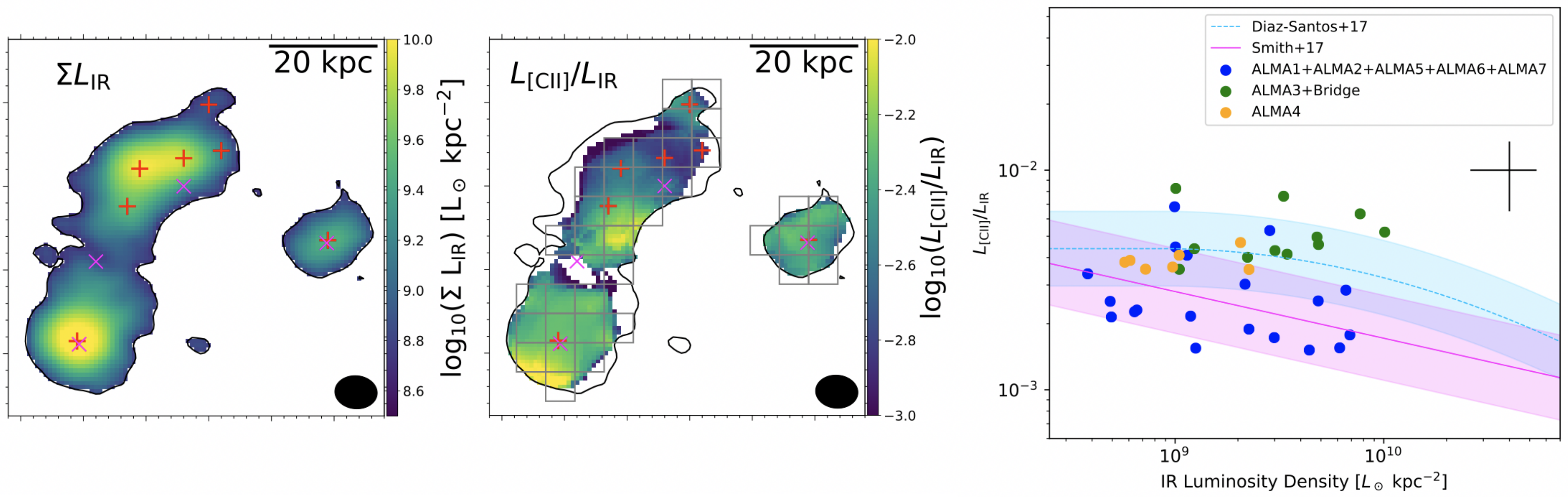}
\caption{
The left panel shows an IR luminosity density map calculated based on the 1.0$^{\prime\prime}$ 860\,$\mu$m map. The middle panel shows the resolved $L_{\rm [CII]}/L_{\rm IR}$ distributions for the areas where both dust and [C~\textsc{ii}] are detected.
The right panel shows $L_{\rm [CII]}/L_{\rm IR}$ as a function of IR luminosity density, using the binned values from the middle panel. 
The [C\,\textsc{ii}]/FIR ratio in LAB1 are broadly consistent with the empirical relations derived by \citet{2017ApJ...834....5S} or the GOALS sample (\citealt{2017ApJ...846...32D}).
}
\label{fig:ciiratio}
\end{figure*}

Both dust and [C\,\textsc{ii}] emission are widely distributed across LAB1, which allows us to investigate the corresponding ISM properties in a spatially resolved manner. 
As shown in Figure~\ref{fig:ciiratio}, a IR luminosity density ($\Sigma L_{\rm IR}$\,[$L_\odot$\,kpc$^{-2}$]) map was calculated on the basis of the 1.0$^{\prime\prime}$ 860\,$\mu$m map as described in Section~\ref{sec:mesurement}.
A $L_{\rm [CII]}/L_{\rm IR}$ map was then constructed by combining with the 1.0$^{\prime\prime}$ [C\,\textsc{ii}] map. 
The right hand panel of Figure~\ref{fig:ciiratio} shows the  distribution of $L_{\rm [CII]}/L_{\rm IR}$ across LAB1 as a function of $\Sigma L_{\rm IR}$.
For comparison, the best-fit function and 1$\sigma$ dispersion derived for local (U)LIRGs (GOALS, \citealt{2017ApJ...846...32D}) and a range of resolved galaxies including local sources as well as high redshift galaxies (\citealt{2017ApJ...834....5S}) are also displayed.

It has been established that $L_{\rm [CII]}/L_{\rm IR}$ decreases as $L_{\rm IR}$ increases (e.g., \citealt{1997ApJ...491L..27M}; \citeyear{2001ApJ...561..766M}; \citealt{2010ApJ...724..957S}). This is sometimes called ``[C\,\textsc{ii}]-deficit''. It has been reported that $L_{\rm [CII]}/L_{\rm IR}$ is more tightly correlated with $\Sigma L_{\rm IR}$ (or the SFR surface density, $\Sigma {\rm SFR}$) (\citealt{2017ApJ...834....5S}; \citealt{2017ApJ...846...32D}). \citet{2017ApJ...846...32D} propose that the radiation field intensity to gas density ratio, which is related to $\Sigma L_{\rm IR}$, is the driver of the [C\,\textsc{ii}]-deficit.
The new ALMA data enable us to construct a resolved picture for the relation in the low IR density regime at $z=3.1$ ($\Sigma L_{\rm IR}\approx5\times10^8-10^{10}$~[L$_\odot$\,kpc$^{-2}$]).
As shown in Figure~\ref{fig:ciiratio}, the $L_{\rm [CII]}/L_{\rm IR}$ ratios in LAB1 are broadly consistent with those observed in local galaxies. This demonstrates that the star-forming galaxies in LAB1 exhibit similar trend to those found in the local Universe. Compared with IR-brighter galaxies at high redshift, they may have moderate radiation field strengths.
Thus the relation between $L_{\rm [CII]}/L_{\rm IR}$ and $\Sigma L_{\rm IR}$ shown in Figure~\ref{fig:ciiratio}
supports the idea that the observed, extended ISM, traced by [C\,\textsc{ii}] and the dust continuum, is mainly associated with star formation: as a part of the massive galaxies, a collection of star-forming galaxies that are individually unresolved, or PDRs in outflows.
We note that there is currently no evidence to support AGN activity in LAB1 (\citealt{2009ApJ...700....1G}), although we cannot exclude the possibility that there is a heavily obscured AGN present. Nevertheless, the observed range of $L_{\rm [CII]}/L_{\rm IR}$ is higher than the value reported for $z=1-2$ AGNs (log($L_{\rm [CII]}/L_{\rm IR}$)$\sim-3.5$ for AGNs with $10^{13}-10^{14}L_{\rm IR}$, \citealt{2010ApJ...724..957S}), which is also consistent with a lack of AGN activity.


One plausible explanation for the presence of [C\,\textsc{ii}] and dust continuum beyond the stellar counterparts may be stripping of highly enriched gas from the less strongly bound regions of galaxies from tidal interactions with one another. In addition to interactions among ALMA-identified galaxies, ALMA3 have two nuclei in the {\it HST} image, which is suggestive of a late-stage merger (Appendix \ref{sec:A.f814w}). The interactions can strip a significant amount of gas and dust out into the inter-group medium. 
The presence of undetected companion galaxies with significant dust extinction could also contribute. Future sensitive imaging at optical-to-near-infrared wavelengths, including {\it JWST}, will improve our understanding in this regard.

There may be another possible origin for the extended [C\,\textsc{ii}] emission: while [C\,\textsc{ii}] emission is expected to arise primarily from PDRs, it is also observed in other environments, including shocked gas (e.g., \citealt{1991ApJ...373..423S}; \citealt{2013ApJ...777...66A}). \citet{2013ApJ...777...66A} reported the detection of [C\,\textsc{ii}] emission associated with shocked gas in a local interacting galaxy group, Stephan's quintet. Such regions show higher luminosity ratios, Log ($L_{\rm [CII]}/L_{\rm IR}) \approx-1.5$ to $-1.0$, which are difficult to explain by star formation alone. The specific environment in LAB1: ongoing mergers and possible association with gas accretion from the CGM/IGM may point to this origin for [C\,\textsc{ii}] emission.
However, as discussed above, the measured $L_{\rm [CII]}/L_{\rm IR}$ appears to be smaller than expected if shocked gas is dominant mechanism for exciting [C\,\textsc{ii}]. Although, the current estimate has relied on SED templates of bright dusty star-forming galaxies due to the shortage of photometry; this may overestimate the dust temperature for an intergalactic region (or a region in the outskirts of a galaxy) and thus underestimate the $L_{\rm [CII]}/L_{\rm IR}$. 
%

In summary, the most plausible scenario is that the observed [C\,\textsc{ii}] emission is principally associated with star formation, although other origins are not excluded. 

\subsection{Galaxy Assembly via multiple-merging}

\begin{figure}
\epsscale{1.15}
\plotone{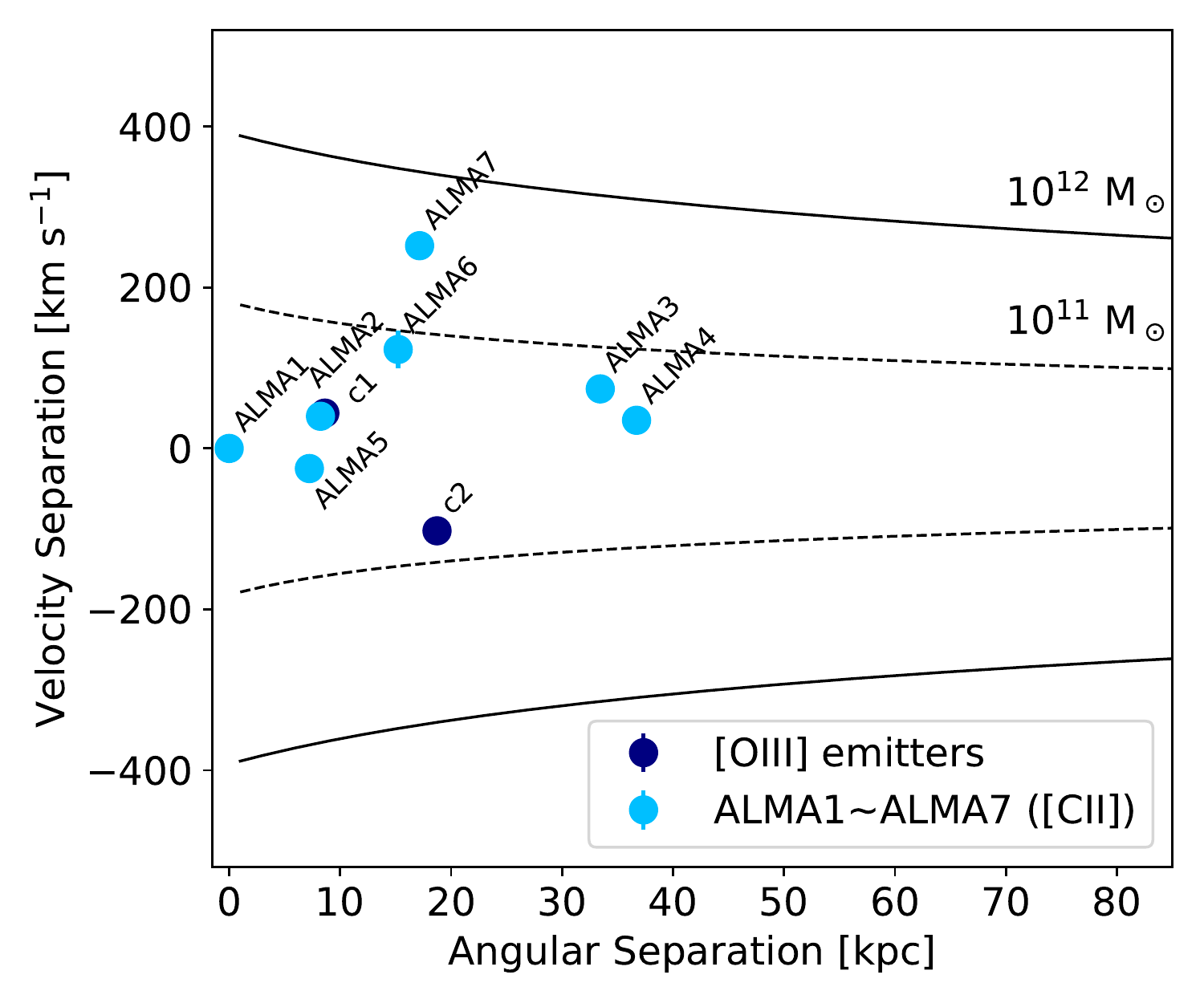}
\caption{
Velocity offsets of the nine sources in LAB1 as a function of projected, physical distance. ALMA1 is assumed to be the center, and navy and light blue points show the [C\,\textsc{ii}]-emitting galaxies and [O\,\textsc{iii}] emitters (\citealt{2021MNRAS.502.2389L}). Tracks show the escape velocities in the case of a NFW halo with a halo mass ($M_{\rm halo}=10^{11}$\,$M_\odot$ and  $10^{12}$\,$M_\odot$). The galaxies in LAB1 have small velocity offsets, which suggests that they are in the merging phase and not virialized yet.
}
\label{fig:phase}
\end{figure}

ALMA observations reveal that LAB1 is associated with a number of star-forming galaxies. In total, seven galaxies are identified from dust continuum and [C\,\textsc{ii}] together with [O\,\textsc{iii}]\,$\lambda$5008 and Ly\,$\alpha$ lines (\citealt{1998ApJ...492..428S}; \citealt{2013ApJ...767...48M}; \citealt{2015ApJ...799...38K};  \citealt{2017ApJ...834L..16U}; \citealt{2021MNRAS.502.2389L}). There are also two additional [O\,\textsc{iii}] emitters (\citealt{2016ApJ...832...37G}; \citealt{2021MNRAS.502.2389L}), 
and thus to date a total of nine galaxies are confirmed to be located in LAB1 within a projected distance of its center of 40\,kpc (here we include  all seven components identified at 860\,$\mu$m, while ALMA6 and ALMA7 may be a tidal tails of ALMA2 as discussed in Section~\ref{sec:cii}). In addition to the overdensity, interactions among multiple galaxies are suggested from [C\,II] (as described in Section \ref{subsec:ciiprop}). Resolved stellar morphologies are also suggestive of some sort of interaction, showing clumps and tails (Section~\ref{sec:cii}). All together, LAB1 appears to be a site of hierarchical galaxy assembly during a star-forming phase in the early Universe. The situation resembles previously reported clusters of bright SMGs, e.g., SSA22-AzTEC14 at $z=3.1$ (\citealt{2015ApJ...815L...8U}; \citealt{2016MNRAS.455.3333K}) and the distant red core (DRC) at $z=4.0$ (\citealt{2018ApJ...856...72O}; \citealt{2020MNRAS.496.4358I}) (see also \citealt{2018Natur.556..469M} and \citealt{2020MNRAS.495.3124H}), although the components of LAB1 are fainter than those in these systems, providing insights into a poorly understood regime.

In order to investigate the state of LAB1 in the context of galaxy assembly, we use clues from the geometry and dynamics of the member galaxies. Figure~\ref{fig:phase} shows the line-of-sight velocity offsets for the $z\approx3.10$ galaxies as a function of projected separation. ALMA1, which is one of the most massive systems in the field and is located close to the centroid of the extended Ly$\alpha$ emission, is adopted as the center. We also plot escape velocities following  \citet{2016MNRAS.455.3333K}. Assuming a Navarro-Frenk-White (NFW) mass profile (\citealt{1997ApJ...490..493N}), the escape velocities for a halo of mass $M_{\rm halo}=10^{11}$\,$M_\odot$ and  $10^{12}$\,$M_\odot$ is calculated as a function of separation from the center. We adopted a concentration parameter $c=4.5$ (\citealt{2011ApJ...740..102K}). In order to take the projection effect into account, the physical distance $r$ and the line-of-sight velocity offset $v$ are corrected using an averaged projection factor $2/\pi$ and $1/\sqrt{3}$, respectively (e.g., \citealt{2015MNRAS.448.1715J}). As shown in Figure~\ref{fig:phase}, the range of velocity separation is relatively small. %
This is in contrast with another dense group in SSA22 (AzTEC14, \citealt{2016MNRAS.455.3333K}), which has a velocity separation $\approx1000$\,km\,s$^{-1}$ with a similar angular separation. 

In LAB1, the galaxies are distributed within a region expected if they are bounded in a halo with $\lesssim10^{12}$\,$M_\odot$.
However, this scenario is highly unlikely. Among the $z\approx3.1$ galaxies, ALMA1, ALMA2, ALMA3, and ALMA4 have $Ks$-band counterparts (Section~\ref{sec:cii}). The sum of the stellar masses of the four galaxies (i.e., the most massive four members) is estimated to be $M_*\approx2.2\times10^{11}$\,$M_\odot$ (e.g., \citealt{2016MNRAS.455.3333K}). Hence the stellar-mass-halo mass relation, which is derived from clustering analysis, suggests a halo with $\sim10^{13}$\,$M_\odot$ (e.g., \citealt{2015A&A...576L...7D}).
We note that a caveat is the influence of line of sight projection on the measured velocity dispersion.
While a simple, spherical geometry is assumed in deriving the track of escape velocity in Figure~\ref{fig:phase}, this may not the case. If we see galaxies in filaments from a viewing angle close to face-on, the apparent velocity offset can be small. Such an effect could account for the observed small velocity range.
Considering the data, therefore one possible scenario is that we are witnessing a merging phase of multiple galaxies hosted in multiple halos. If they are merging and not yet virialized, the velocity offsets are expected to be small as observed in LAB1. A multiple-merger phase of star-forming galaxies is expected for a progenitor of a brightest cluster galaxy (BCG) in the nearby Universe during hiearchical galaxy assembly in the early Universe (e.g., \citealt{2016MNRAS.455.3333K}), and thus an evolutional connection between BCG formation and LABs may be suggested. 
This phase tends to occur in halo masses of group size, consistent with our observations.

One issue to be resolved regarding the assembly phase is the deficit of observed low-mass galaxies (e.g., \citealt{2009MNRAS.395..114H}; \citealt{2016MNRAS.455.3333K}). For instance, \citet{2009MNRAS.395..114H} investigated the stellar mass distribution of the spiderweb proto-cluster at $z=2.2$, and reported that the number of observed galaxies falls for short at
 $M_*<10^{9.8}$~$M_\odot$, compared to the prediction of semi-analytic models. The existing {\it K}-band image of LAB1 allows us to detect relatively unobscured galaxies with $M_*\gtrsim10^{10.5}$\,M$_\odot$ at $z\sim3$ (\citealt{2016MNRAS.455.3333K}).
 Galaxies newly discovered in [C\,\textsc{ii}] may account for (some of) these galaxies which are ISM-rich and had been evaded for detection at UV-to-optical wavelengths due to the extinction. Sensitive census of stellar mass distributions, which will be possible with the {\it JWST}, will allow us to further test such a scenario.

\subsection{The origins of Ly$\alpha$ emission and baryon cycling} \label{subsec:lyaorigin}

\begin{figure*}
\epsscale{1.15}
\plotone{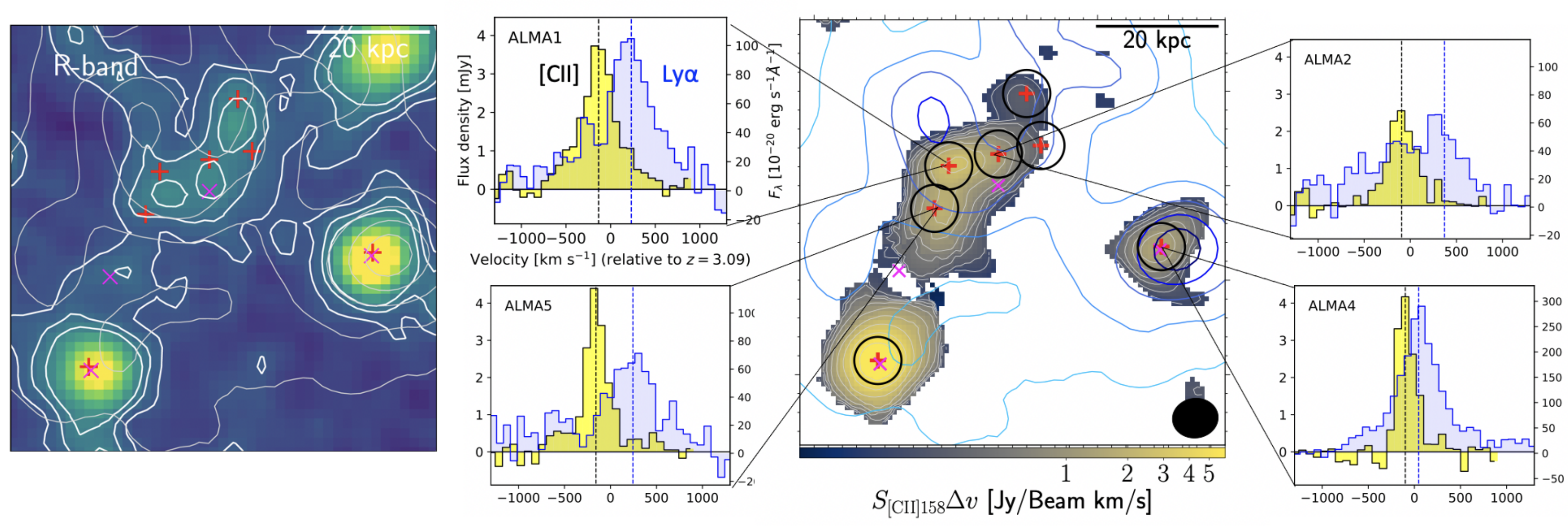}
\caption{
The left panel shows a Subaru/Suprime-Cam {\it R}-band image, which traces rest-frame UV ($\approx$ 1550\,\AA) emission at $z\approx3.1$. Thick contours show [3, 6, 9]$\times \sigma$ per arcsec$^2$ (\citealt{2007ApJ...667..667M}). 
Red and magenta crosses shows ALMA sources and [O\,III]-emitting galaxies as previous figures.
The right hand image shows the rendered, velocity-integrated [C~\textsc{ii}] intensity map at 1.0$^{\prime\prime}$ resolution with contour levels of $0.1 \times [1.5^2, 1.5^3,...]$~Jy~beam$^{-1}$ km~s$^{-1}$. Superposed contours with blue colors show the Ly\,$\alpha$ emission as in Figure~\ref{fig:cont1}. Both [C\,\textsc{ii}] and Ly\,$\alpha$ spectra extracted with a 1$^{\prime\prime}$ diameter aperture are also presented for four ALMA components identified at 860\,$\mu$m.
The Ly\,$\alpha$ emission generally has a dominant peak that is redshifted with respect to [C\,\textsc{ii}], which indicates either outflow motion of the H\,\textsc{i} gas or absorption around the ALMA components.}
\label{fig:cii_tail}
\end{figure*}

The physical nature of Ly\,$\alpha$ blobs is still without a consensus model, with several competing theories.
What mechanisms power the enormous and extended Ly\,$\alpha$ emission is a key issue. The kinematics of the neutral hydrogen traced by Ly\,$\alpha$ (i.e.,, inflow and outflow) is also of interest in understanding baryon cycling in a massive halo. LAB1 has been a remarkable target in these contexts since its discovery (e.g., \citealt{2000ApJ...532..170S};  \citealt{2004MNRAS.351...63B}; 
\citealt{2004AJ....128..569M};
\citealt{2004ApJ...606...85C};
\citeyear{2007ApJ...667..667M};
\citealt{2005MNRAS.363.1398G}; \citeyear{2009ApJ...700....1G}; \citeyear{2014ApJ...793...22G}; \citeyear{2016ApJ...832...37G}; \citealt{2006Natur.440..644M}; 
\citealt{2010MNRAS.402.2245W}; \citealt{2011Natur.476..304H};
\citealt{2013ApJ...775..112C}; 
\citealt{2013MNRAS.430.2768T}; \citealt{2013ApJ...773..151Y}; \citealt{2016MNRAS.460.4075H}; \citealt{2017ApJ...850..178A};
\citealt{2017ApJ...834L..16U};
\citealt{2020A&A...642A..55H}; 
\citealt{2021MNRAS.502.2389L}). Our new ALMA observations provide key information for reassessing LAB1's nature.

\subsubsection{The central heating source and scattering scenario}

One scenario proposed for the source that powers the Ly\,$\alpha$ emission is the existence of a central heating source which provides ionizing photons via star-formation or AGN (e.g., \citealt{2000ApJ...532..170S}; \citealt{2009ApJ...700....1G}). While AGNs are not confirmed in LAB1 on the basis of X-ray (\citealt{2009ApJ...700....1G}) or radio observations (\citealt{2017ApJ...850..178A}),
several galaxies which harbor dust-enshrouded  star formation have been identified (e.g.,
\citealt{2004ApJ...606...85C};
\citealt{2005MNRAS.363.1398G}; \citealt{2013MNRAS.430.2768T}; \citealt{2014ApJ...793...22G}).
Most recently, \citet{2016ApJ...832...37G} identified three dusty sources, corresponding to ALMA1, ALMA2, and ALMA3 in this paper. They showed that Ly\,$\alpha$ photons escaping from these dusty sources could generate the bright, extended Ly\,$\alpha$ emission as a result of successive scattering, using a cosmological zoom-in simulation. This is also in line with the detection of a polarized ring nearly centered at the Ly$\alpha$ emission peak (and the position of ALMA1, \citealt{2011Natur.476..304H}). \citet{2021MNRAS.502.2389L} also reported that the observed Ly\,$\alpha$/H\,$\beta$ ratios are explainable by this scenario, although H\,$\beta$ fluxes detected by the authors are generally only from galaxies and there is still room for discussion about the extended Ly\,$\alpha$ emission.

Here we start with the ``central source(s) and scattering'' scenario making use of what we have learned from the ALMA data. One concern in \citet{2016ApJ...832...37G} was the absence of $z_{\rm spec}$ for the two dusty star-forming galaxies located near the center of LAB1. The line detections ([C\,\textsc{ii}], CO(4-3)) from ALMA1+2+5 definitively show that the dusty star-forming galaxies are physically associated with LAB1. On the basis of the new 860\,$\mu$m map, the integrated 860\,$\mu$m flux $S_{860}\approx2.4$~mJy implies SFR$_{\rm IR}\approx210$~M$_\odot$~yr$^{-1}$ in total. 

In addition, the line detections shed light on the faint, relatively {\it unobscured} star-formation. As \citet{2007ApJ...667..667M} reported, there is extended emission in {\it R}-band within LAB1, which encompasses the ALMA-identified galaxies (Figure~\ref{fig:cii_tail}). The detected [C\,\textsc{ii}] emission is coincident with the emission in {\it R}-band and thus likely to be rest-frame UV (centered at around 1550\AA) emission at $z=3.1$. We measured the total UV flux in regions enclosed by the 3$\sigma$ contours in {\it R}-band and located within the apertures that enclose the [C\,\textsc{ii}] emission (ALMA1+2+5, ALMA3, and ALMA4), $F_\nu\approx1.3\times10^{30}$~erg~s$^{-1}$~Hz$^{-1}$. The inferred SFR from this is SFR$_{\rm UV} \approx90$~M$_\odot$~yr$^{-1}$ following the equation SFR$_{\rm UV}$~[M$_\odot$~yr$^{-1}$]$=L_\nu \times (1.46\times10^{21})^{-1}$~[W~Hz$^{-1}$] (for the Chabrier IMF, \citealt{1998ARA&A..36..189K}; \citealt{2007ApJS..173..267S}), and so total UV+IR SFR is $\sim$ 300\,M$_\odot$\,yr$^{-1}$.
The three apertures show roughly equivalent contribution, so that the obscured fractions of total SFR in ALMA1+2+5, ALMA3, and ALMA4 are $\sim80\%$, $\sim80\%$, and $\sim40\%$, respectively.

This star forming activity causes Ly\,$\alpha$ emission and also provide the ionizing photons ($\nu$=200-912\AA) to the surrounding environment (if they escape).
Under the assumption of Case-B recombination, the Ly\,$\alpha$ luminosity generated by the star-formation that escape the host galaxy are described as follows.
\begin{equation}
L_{\rm Ly\alpha} [{\rm erg}~{\rm s}^{-1}] = f_{\rm esc, Ly\alpha} \times 1.89\times10^{42}\times {\rm SFR} [M_\odot {\rm yr}^{-1}]
\end{equation}
Here $f_{\rm esc, Ly\alpha}$ is the escape fraction of Ly$\alpha$ photons. 
%
Considering the Ly\,$\alpha$ luminosity of LAB1, L$_{\rm Ly\alpha}\approx10^{44}$~erg~s$^{-1}$, therefore the star-forming activities can power the Ly\,$\alpha$ emission if the escape fractions are high enough ($f_{\rm esc, Ly\alpha}\gtrsim20\%$).

In this scenario, Ly\,$\alpha$ photons, produced  by star-forming galaxies scatter through circumgalactic  H\,\textsc{i}, producing an extended ``Ly\,$\alpha$ halos'' (e.g., \citealt{2011ApJ...736..160S}).
Thus the gas traced by Ly\,$\alpha$ emission is expected to associate with outflow motion. As shown in Figure~\ref{fig:cii_tail}, the profiles of Ly$\alpha$ spectra have a common feature at the positions of ALMA components: Ly$\alpha$ has a minimum near the systemic redshift measured by [C\,\textsc{ii}], with a clear redshifted dominant peak and a blueshifted component.
This trend is inline with the idea that the observed Ly\,$\alpha$ emission (mostly) comes from scattering from outflowing gas; outflowing Ly\,$\alpha$ photons that are back-scattered from gas on the far are the most likely to escape obtaining a frequency shift, while absorption by neutral hydrogen and dust is the most significant at the galaxy's redshift (e.g., \citealt{2011ApJ...736..160S};  \citealt{2015Natur.523..169E}; \citealt{2020arXiv201203959C}). Outflows may also contribute to increasing the covering fraction of neutral hydrogen (\citealt{2015MNRAS.452.2034R}).

Recently, \citet{2021MNRAS.502.2389L} performed a Monte-Carlo radiative transfer modeling for LAB1 using their Ly\,$\alpha$ cube taken with KCWI. They reproduce the observed Ly\,$\alpha$ spectra and constrain parameters in their outflow model with multi-phase and clumpy hydrogen gas. They reported that a region around ALMA1 has the highest optical depth and H\,\textsc{i} outflow velocity in the ionized inter-clump medium. ALMA1 is actually at the redshift inferred by their model and  the starburst galaxy is likely responsible for these feature, as the authors predict.

Thus this central heating and scattering scenario is further supported by the newly delivered ALMA data in some respects. The dusty star-forming galaxies uncovered by ALMA undoubtedly play a significant role 
in powering Ly\,$\alpha$ emission and ejecting gas, metals, and dust into the surrounding medium.

\subsubsection{Additional sources to power Ly\,$\alpha$ emission}

While star formation can apparently explain the Ly\,$\alpha$ properties of LAB1, it is still unclear whether it is the solo mechanism causing the enormous Ly\,$\alpha$ emission. While the polarized ring discovered by \citet{2011Natur.476..304H} indicates a heating source  near the Ly\,$\alpha$ emission peak, only half of the total SFR comes from the central group. The remaining fractions evidently originate from the outskirts. Furthermore the obscured fraction of SFR for ALMA1+2+5 and ALMA3 is $\sim80$\%, and it is uncertain whether these regions would be expected to have relatively high escape fractions of Ly\,$\alpha$ photons.
There is another clue to strengthen such a caveat. As reported in \citet{2017ApJ...835...98U} and \citet{2018PASJ...70...65U} (see also \citealt{2015ApJ...815L...8U}), there are a number of SMGs at $z\approx3.09$ which show higher levels of star-forming activity in the SSA22 proto-cluster. While a sensitive census of extended Ly\,$\alpha$ emission shows that the SMGs ubiquitously reside in Ly\,$\alpha$ filaments, the associated Ly\,$\alpha$ emission is usually fainter than LABs, in contrast to the activity of associated star formation (\citealt{2019Sci...366...97U}; other examples in \citealt{2018ApJ...856...72O}; \citealt{2019ApJ...875..130L}). 
There is also no evidence favoring exceptionally high escape fractions just for the galaxies in LAB1.
These results suggest that the total SFR associated with Ly\,$\alpha$ nebulae is not the only key parameter for diagnosing the powering sources of the observed Ly\,$\alpha$ emission. 

The exceptionally bright and extended Ly\,$\alpha$ nature of LAB1 therefore suggests that there is additional mechanism at play especially in LAB1.
A possible path to explain the enhanced Ly\,$\alpha$ level in LAB1 is proposed by \citet{2013ApJ...773..151Y}. On the basis of a combination of hydrodynamical simulations with three-dimensional radiative transfer calculations, they investigated the environment surrounding a galaxy-galaxy major-merger. They successfully produced mock LABs with luminosity of $L_{\rm Ly\alpha}\sim10^{42}$--10$^{44}$\,erg\,s$^{-1}$ and extent of $\sim50$~kpc at $z\sim3$. The authors find that both merger-driven intense star formation and cooling radiation induced by strong gravitational interactions contribute to generate Ly\,$\alpha$, although the relative fractions are not discussed. While the simulation is only for a binary major-merger, they suggest that multiple-mergers can generate more spectacular nebulae. As discussed above, LAB1 is found to host ongoing multiple-mergers close to the Ly\,$\alpha$ emission peak. This is a unique characteristic of LAB1, and suggestive that shocked (collisionally excited) Ly\,$\alpha$ may need to be included.

In addition, cold accretion along the filaments has been suggested to be a powering source for years (e.g., \citealt{2009MNRAS.400.1109D}). \citet{2016A&A...593A.122T} used a radiative hydrodynamics simulation and showed that the observed polarization is explainable with the combination of the `central powering $+$ scattering' model and Ly$\alpha$ emission originated from gas during the accretion onto the halo. There is the dominant molecular gas reservoir close to the Ly\,$\alpha$ emission peak, which is indicative of gas supply from the cosmic web onto the center of LAB1. Therefore the ALMA data is not necessarily conflict with the scenario of the cold accretion in this sense.




\section{Conclusions} \label{sec:conclusion}

We have carried out deep multi-band observations of the giant Ly\,$\alpha$ nebulae SSA22-LAB1 at $z=3.1$ using ALMA.
The main conclusions are the following:

\begin{itemize}

\item[1.] We performed the most sensitive census to date of dust continuum emission in LAB1 at observed wavelengths of 656\,$\mu$m, 860~$\mu$m, and 2.82\,mm. The 860~$\mu$m maps uncover an extended structure of dust emission on a 60\,kpc scale, which is decomposed to eight individual components (ALMA1--ALMA7, and the Bridge). 

\item[2.] [C\,\textsc{ii}]\,158$\mu$m emission is also widely distributed across LAB1, generally coincident with the 860~$\mu$m dust continuum. Moment maps and P--V diagrams suggest possible ongoing multiple-merging events involving three dusty galaxies, ALMA1, ALMA2, and ALMA5. 

\item[3.] Comparison with optical-to-infrared images demonstrates that our previous view of the components of LAB1 has been biased due to heavy dust extinction. A remarkable example is ALMA5, which has no counterpart at near-infrared or optical wavelengths, while a rotating disk structure is suggested by its [C\,\textsc{ii}] emission. The ALMA census sheds light on a number of previously missed LAB1 members.

\item[4.] Massive molecular gas reservoirs with $M_{\rm gas}\approx10^{11}$\,M$_\odot$ are uncovered in LAB1 from CO(4--3) emission. We found that the majority of this gas mass is concentrated near the Ly\,$\alpha$ peak.



\item[5.] The components of LAB1 identified in [C\,\textsc{ii}] and [O\,\textsc{iii}]\,$\lambda$5008 show a tight range in redshift, $z=3.0968-3.1016$. 
LAB1 seems to be multiple-merging phase involving a number of galaxies and halos on a group scale, which may be a progenitor of a Bright Cluster Galaxy.

\item[6.] The derived (UV+IR) SFRs and profiles of Ly\,$\alpha$ spectra around dusty galaxies suggests their important role in powering the extended Ly\,$\alpha$ emission as a heating source. However, it is not clear whether or not star formation in the galaxies in solely responsible. We suggest that cooling radiation induced by strong gravitational interactions may also play a significant role.

\end{itemize}


\acknowledgments
We wish to thank the anonymous referee for constructive comments that improved this paper.
%
We acknowledges valuable discussion with Hidenobu Yajima. We thank Scott Chapman and Ken Mawatari for sharing calibrated optical-to-near-infrared images with us. 
H.U. and Y.M. acknowledge support from JSPS KAKENHI grant (17KK0098, 20H01953).
I.R.S. acknowledges STFC through grant number ST/T000244/1.
R.J.I is funded by the Deutsche Forschungsgemeinschaft (DFG, German Research Foundation) under Germany's Excellence Strategy -- EXC-2094 --390783311.
This paper makes use of the following ALMA data: ADS/JAO.ALMA \#2013.1.00704.S, \#2013.1.00922.S, \#2016.1.00485.S, \#2016.1.01134.S,  \#2017.1.01209.S. ALMA is a partnership of ESO (representing
its member states), NSF (USA) and NINS (Japan), together with NRC
(Canada), MOST and ASIAA (Taiwan), and KASI (Republic of Korea), in cooperation with the Republic of Chile. The Joint ALMA Observatory is operated by ESO, AUI/NRAO and NAOJ. The National Radio Astronomy Observatory is a facility of the National Science Foundation operated under cooperative agreement by Associated Universities, Inc. 
Our data are based on
observations collected at the European Organisation for
Astronomical Research in the Southern Hemisphere.
This research is based on data collected at Subaru Telescope, which is operated by the National Astronomical Observatory of Japan. We are honored and grateful for the opportunity of observing the Universe from Maunakea, which has the cultural, historical and natural significance in Hawaii.
This research is based on observations made with the NASA/ESA Hubble Space Telescope obtained from the Space Telescope Science Institute, which is operated by the Association of Universities for Research in Astronomy, Inc., under NASA contract NAS 5--26555. 

\appendix

\section{The 860\,$\mu$m image and source decomposition}\label{sec.A1}

\begin{figure*}
\epsscale{1.05}
\plotone{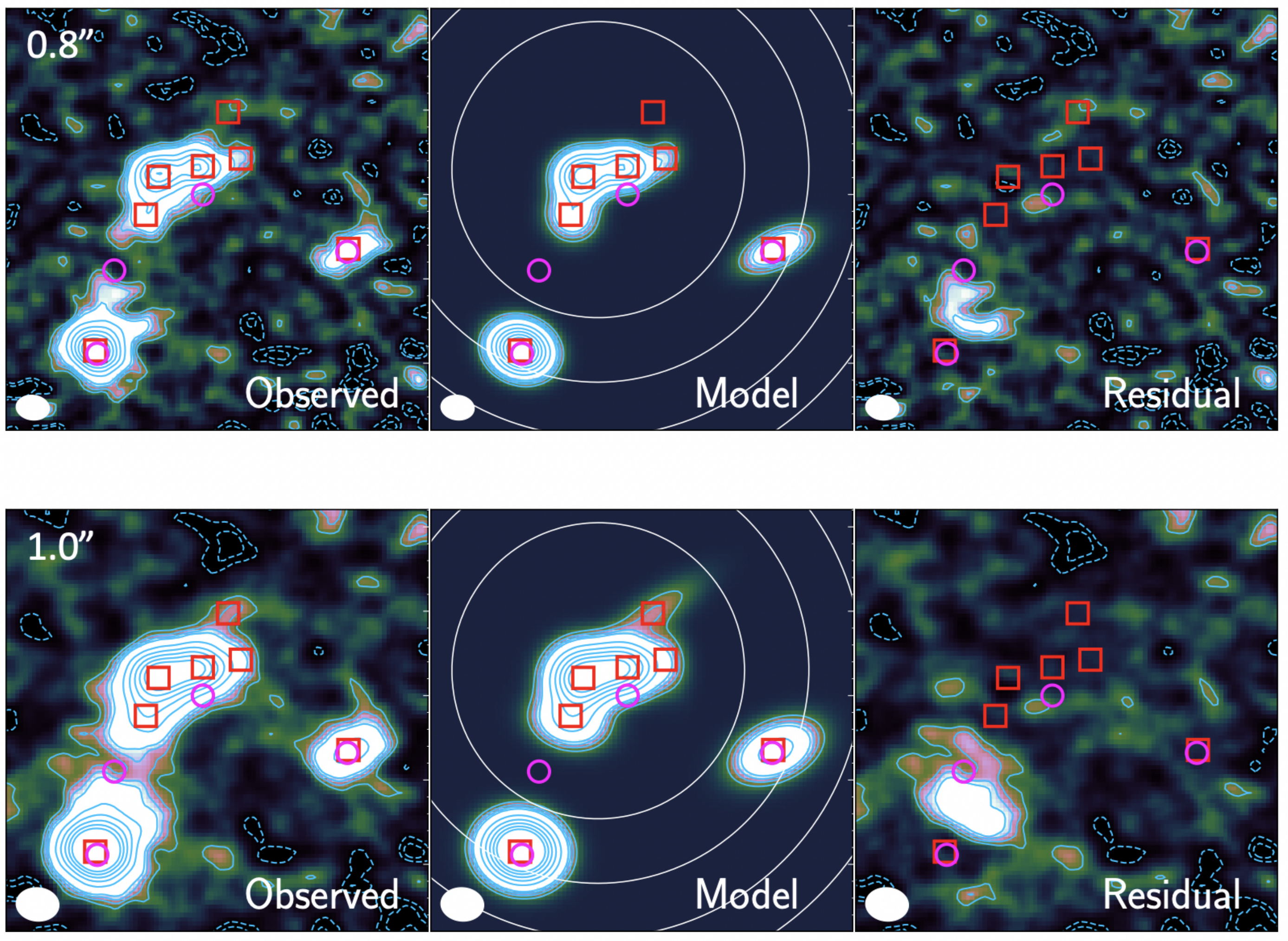}
\caption{
(Left) ALMA 860\,$\mu$m image at two angular resolutions ($0.8^{\prime\prime}$ and $1.0^{\prime\prime}$ as labelled) as shown in the bottom middle panel of Figure~\ref{fig:cont1}. Note that the map after primary beam correction is shown here. Contours shows [-3,-2,2,3,4,5,10,15,20,25,30]$\times \sigma_{\rm center}$ emission ($\sigma_{\rm center}$ shows the rms noise level at the phase center). Red squares show positions of $z=3.1$ galaxies identified by ALMA, while magenta circles show those of [O~\textsc{iii}] emitters.
(middle) modelled source profile of six dusty star-forming galaxies at $z=3.1$. Primary beam responses are also shown using white lines, which correspond to 60\,\%, 70\,\%,80\,\%, and 90\,\% of the primary beam response.
(right) residual map after subtracted the model images. Extended emission between ALMA3 and ALMA5 is securely detected in both maps.}
\label{fig:decomp}
\end{figure*}

The 860\,$\mu$m image was utilized to isolate individual components which were closely located each other. First the brightest four galaxies identified in the 656\,$\mu$m image were fitted and subtracted on the image plane using {\sc casa}/{\sc imfit}, and the remaining two galaxies were also similarly fitted. Figure~\ref{fig:decomp} shows the original image, best-fit models, and the residual image. As shown, in total seven individual galaxies are identified at 860\,$\mu$m. 

\section{Channel Maps}

\begin{figure*}
\epsscale{1.15}
\plotone{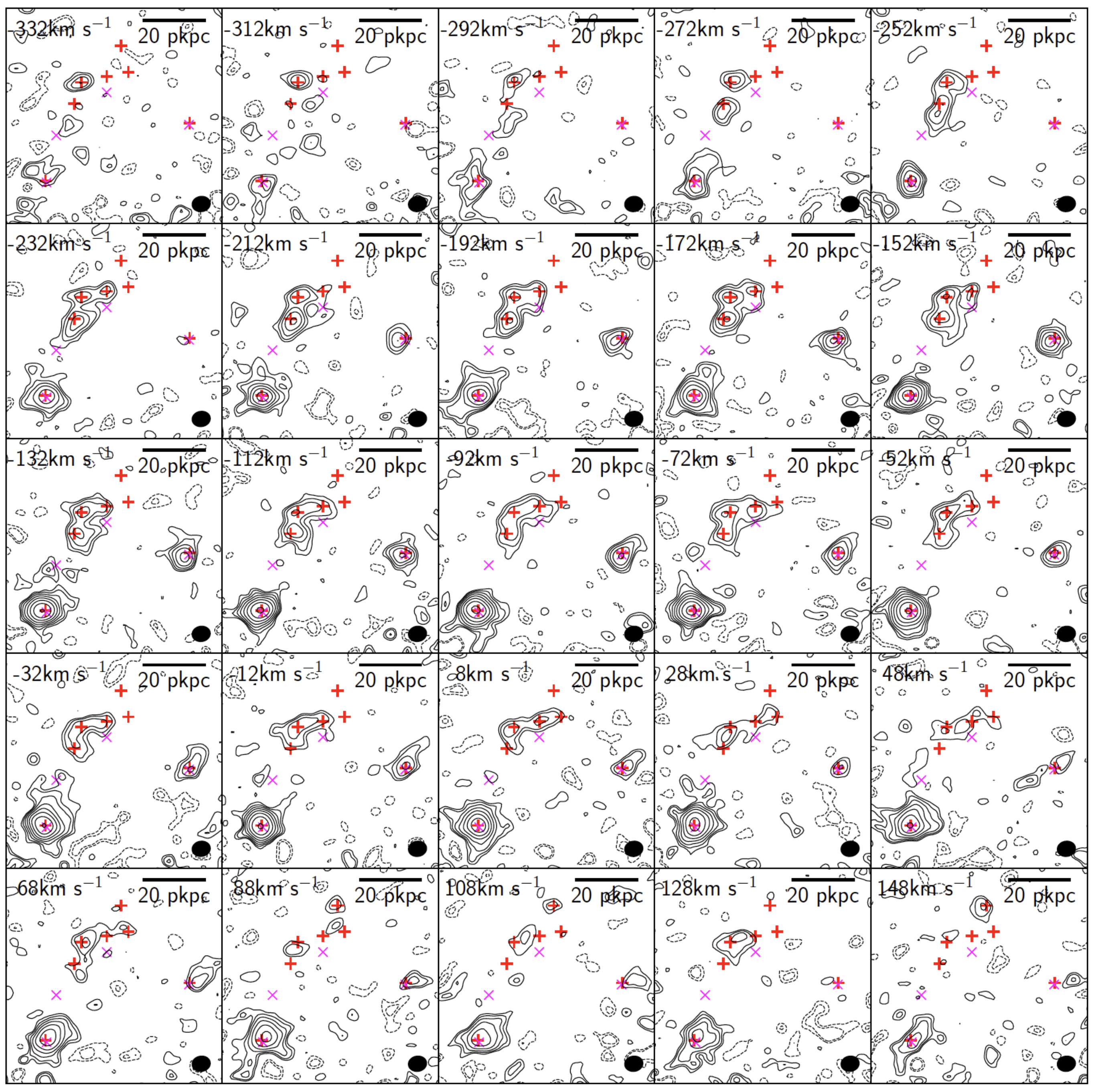}
\caption{
Channel map of [C\,\textsc{ii}] emission in LAB1 in a range of velocity that contains a large fraction of the emission.
 Each panel is $9^{\prime\prime}\times9^{\prime\prime}$ in size (corresponding 70\,kpc$\times$70\,kpc at $z\approx3.1$).
Contours show [ -$1.5^3, -1.5^2, 1.5^2, 1.5^3,...$]$\times \sigma_{\rm center}$ of  [C\,\textsc{ii}] emission (a fixed value of $\sigma_{\rm center}=0.50$\,$\mu$Jy is adopted). 
Crosses show the positions of  860\,$\mu$m-identified components and [O\,\textsc{iii}]~$\lambda 5008$ emitters as Figure~\ref{fig:cont1}.
[C\,\textsc{ii}] emission is distributed across the field with a complex morphology. %
All seven components identified in dust continuum are also detected in [C\,\textsc{ii}] and the emission from dust and [C\,\textsc{ii}] are generally co-spatial across the field.
}
\label{fig:chan1}
\end{figure*}

Figure~\ref{fig:chan1} shows a channel map of [C\,\textsc{ii}] emission in LAB1.

\section{[CII] Spectra of two [OIII] emitters}

\begin{figure*}
\epsscale{1.15}
\plotone{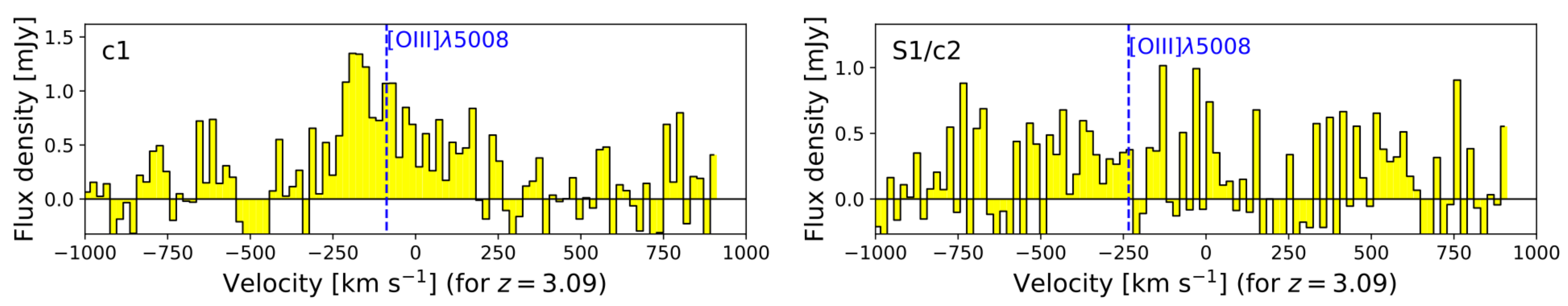}
\caption{
[C\,\textsc{ii}] Spectra of c1 and S1/c2 extracted using a $d=0.8^{\prime\prime}$ aperture.
}
\label{fig:oiiiEspec}
\end{figure*}

Two [O\,\textsc{iii}] emitters were discovered by \citet{2016ApJ...832...37G} and \citealt{2021MNRAS.502.2389L}. The spectra at the reported positions are shown in Figure~\ref{fig:oiiiEspec}. Both are likely to be also associated with [C\,\textsc{ii}] emission, but the blending of the nearby bright [C\,\textsc{ii}] sources, and/or a relatively low signal-to-noise ratio, prevent us from isolating their contributions.

\section{A galaxy previously reported to be at $z=3.1$}

\citet{2015ApJ...799...38K} reported that one $K_{\rm s}$-band-selected galaxy (K1 or K15b) has a [O~\textsc{iii}]~$\lambda$5008 emission at $z=3.1007$ with a moderate significance. However, \citet{2020arXiv200809130L} reported a non-detection of any line from K1 based on their more sensitive observation and hence there is no secure line detection on this source so far. 
\citet{2004ApJ...606...85C} argued that the galaxy is likely at a much lower redshift on the basis of $U, g, R, I$, and $K$ colors.
We cannot detect any [C~\textsc{ii}] (and dust continuum emission) from K1 and so our observations also do not support the idea that K1 is a member of LAB1. We do not include this source in discussion in this paper.
Assuming a 300~km~s$^{-1}$ velocity width, 3$\sigma$ point source limit of K1 is derived to be $S \Delta v=0.2$~Jy~km~s$^{-1}$. For reference, this corresponds to [C~\textsc{ii}] line luminosity limit $L_{\rm [CII]}<7\times10^7$~L$_\odot$ in the case of $z=3.10$.

\section{ISM properties of seven ALMA galaxies in LAB1 at 1.0$\prime\prime$ resolution}\label{sec.A.ciitable}

The measured ISM properties of seven ALMA galaxies at 1.0$^{\prime\prime}$ are summarized in Table~\ref{table:ciiratio}.

\begin{deluxetable*}{ccccccccc}
\tabletypesize{\scriptsize}
\tablecaption{ISM properties LAB1 at a 1.0$^{\prime\prime}$ resolution}
\tablewidth{0pt}
\tablehead{
\colhead{ID} & \colhead{$z_{\rm [CII]}$} & \colhead{$\Delta v$}$^a$  & \colhead{$\Delta v_{\rm Ly\alpha}$}$^b$ & \colhead{FWHM} & \colhead{$S\Delta v$} & \colhead{$L_{\rm [CII]}$} &  \colhead{$S_{\rm 860 \mu m}$ } & \colhead{$L_{\rm IR}$} \\
\colhead{} & \colhead{} &  \colhead{[km~s$^{-1}$]} & \colhead{[km~s$^{-1}$]} & \colhead{[km~s$^{-1}$]}  & \colhead{[Jy~km~s$^{-1}$]} & \colhead{[10$^{8}$~$L_\odot$]} &  \colhead{[$\mu$Jy]} &  \colhead{[10$^{11}$~$L_\odot$]}
}
\startdata
ALMA1   & 3.0982 & -133 $\pm$ 10 & 366 $\pm$ 13 & 482 $\pm$ 25 & 1.75 $\pm$ 0.08 & 6.1 $\pm$ 0.3 & 296 $\pm$ 6  & 2.5 $\pm$ 0.8 \\
ALMA2   & 3.0987 &  -93 $\pm$  8 & 464 $\pm$ 12 & 359 $\pm$ 21 & 0.94 $\pm$ 0.05 & 3.3 $\pm$ 0.2 & 224 $\pm$ 6  & 1.9$\pm$ 0.6 \\
ALMA3   & 3.0992 &  -59 $\pm$  3 & 493 $\pm$ 32 & 303 $\pm$  7 & 5.14 $\pm$ 0.11 & 17.9$\pm$ 0.4 & 407 $\pm$ 8  & 3.4 $\pm$ 1.0 \\
ALMA4   & 3.0986 &  -98 $\pm$  6 & 155 $\pm$  7 & 220 $\pm$ 16 & 0.99 $\pm$ 0.06 & 3.5 $\pm$ 0.2 &  93 $\pm$ 8  & 0.8 $\pm$ 0.2 \\
ALMA5   & 3.0978 & -158 $\pm$  7 & 402 $\pm$ 10 & 274 $\pm$ 18 & 1.21 $\pm$ 0.07 & 4.2$ \pm$ 0.2 & 127 $\pm$ 6  & 1.1 $\pm$ 0.3 \\
ALMA6   & 3.0999 &  -10 $\pm$ 21 & 286 $\pm$ 24 & 241 $\pm$ 51 & 0.23 $\pm$ 0.04 & 0.8 $\pm$ 0.1 &  57 $\pm$ 6  & 0.5 $\pm$ 0.2 \\
ALMA7   & 3.1016 &  119 $\pm$ 11 & 146 $\pm$ 14 & 164 $\pm$ 27 & 0.21 $\pm$ 0.03 & 0.7 $\pm$ 0.1 &  24 $\pm$ 7  & 0.2 $\pm$ 0.1 \\
\enddata
\tablecomments{
Line properties are measured through a Gaussian fit to the spectra with a $d=1^{\prime\prime}$ aperture (Figure~\ref{fig:cii_tail}). Continuum fluxes are also measured using the same apertures and IR fluxes are calculated using a set of SED templates (\citealt{2017ApJ...835...98U}).
$^a$ [C~\textsc{ii}] velocity relative to $z=3.100$.
$^b$ A red peak of Ly\,$\alpha$ velocity relative to [C~\textsc{ii}] velocity.
}
\label{table:ciiratio}
\end{deluxetable*}

\section{ACS image of ALMA3}\label{sec:A.f814w}

\begin{figure*}
\epsscale{0.6}
\plotone{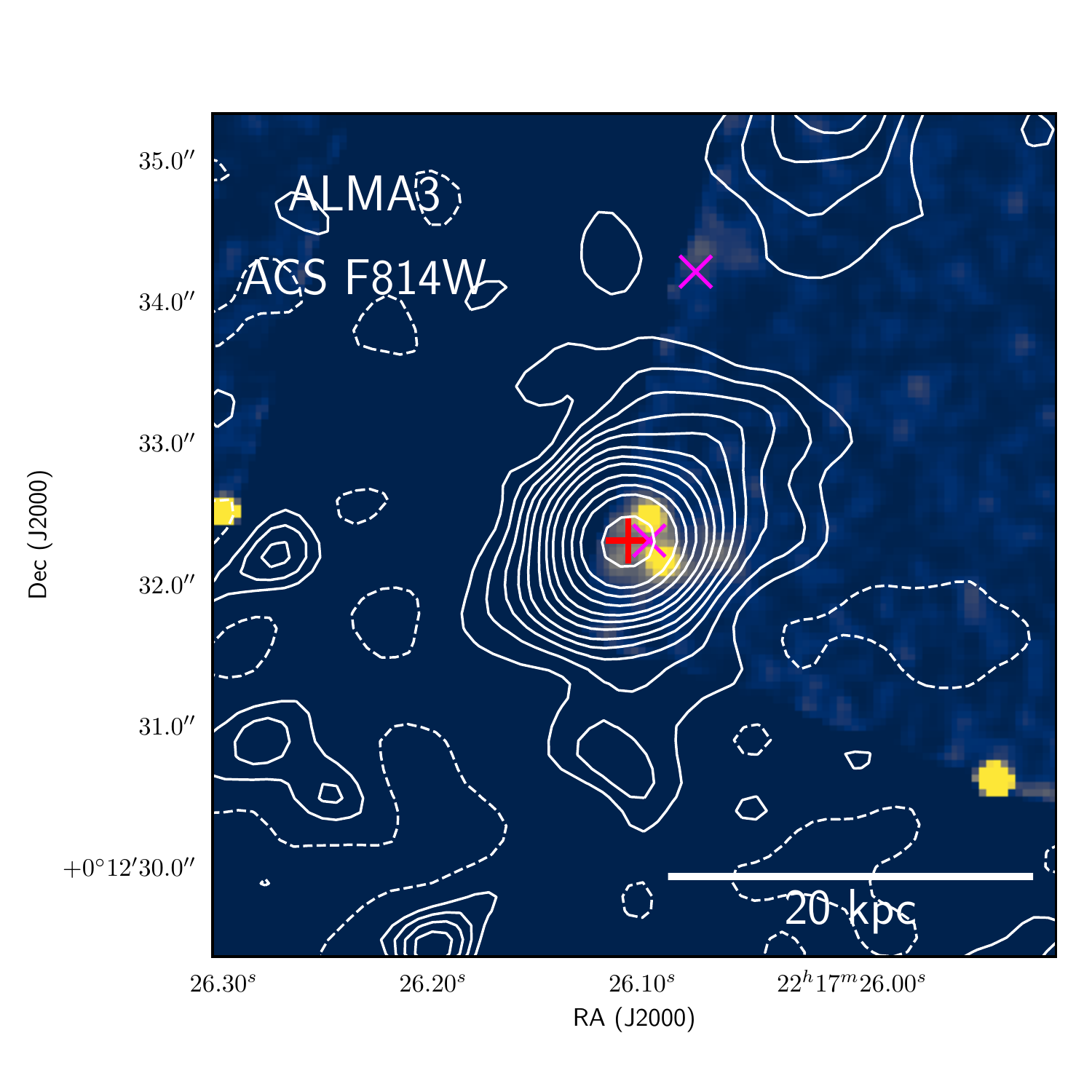}
\caption{
An ALMA3 image taken with {\it HST}/ACS
 using the F814W filter. Contours and markers show [C~\textsc{ii}] emission and galaxy positions as Figure~\ref{fig:hst}. ALMA3 is found to have two nuclei or clumps in the band, which is suggestive of a galaxy-galaxy merger-merger.
}
\label{fig:acs}
\end{figure*}

A {\it HST} image taken with the Advanced Camera for Surveys, ACS, with the F814W filter covers LAB1, while LAB1 is located in the edge region of the map. As shown in Figure~\ref{fig:acs}, ALMA3 is found to have two major-components at the core. This indicates a merger, while other scenarios like dusty lane are not excluded.

%

\vspace{5mm}
\facilities{ALMA, Subaru, VLT(MUSE), HST(STIS)}

\software{
astropy \citep{2013A&A...558A..33A};
casa \citep{2007ASPC..376..127M};
MUSE pipeline \citep{2016ascl.soft10004W};
mpdaf \citep{2017arXiv171003554P};  \citep{2016ascl.soft11003B}
}








\end{document}